\documentclass[electronics,article,accept,moreauthors,pdftex,10pt,a4paper]{mdpi} 

\firstpage{1} 
\articlenumber{57}
\doinum{10.3390/electronics6030057}
\pubvolume{6}
\pubyear{2017}
\copyrightyear{2017}
\history{Received: 22 July 2017; Accepted: 8 August 2017; Published: 10 August 2017}

\usepackage{algorithmic}
\usepackage{array}
\usepackage{mdwmath}
\usepackage{mdwtab}
\usepackage{eqparbox}
\usepackage[tight,footnotesize]{subfigure}
\usepackage{stfloats}
\usepackage{url}
\usepackage{graphicx}
\usepackage{color}
\usepackage{placeins}
\usepackage{float}
\usepackage{cases}
\usepackage{tabularx,colortbl}
\usepackage{steinmetz}
\usepackage{flushend}
\usepackage{amsmath,amsthm,amssymb,mathtools}
\setitemize{parsep=6pt,itemsep=0pt,leftmargin=*,labelsep=5.5mm}
\setenumerate{parsep=6pt,itemsep=0pt,leftmargin=*,labelsep=5.5mm}
\setlist[description]{itemsep=0mm}

\Title{Antenna Arrays for Line-Of-Sight Massive MIMO: Half Wavelength Is Not Enough}

\Author{{Daniele Pinchera} *, Marco Donald Migliore, Fulvio Schettino and Gaetano Panariello}
\AuthorNames{Daniele Pinchera, Marco Donald Migliore, Fulvio Schettino and Gaetano Panariello}

\address[1]{{DIEI and ELEDIA@UNICAS}, University of Cassino and Southern Lazio, via G. Di Biasio 43, 03043 Cassino, Italy; mdmiglio@unicas.it (M.D.M.); schettino@unicas.it (F.S.); panariello@unicas.it (G.P.)}

\corres{\hangafter=1 \hangindent=1.05em \hspace{-0.82em} Correspondence: pinchera@unicas.it; Tel.: +39-0776-299-4348}

\abstract{The aim of this paper is to analyze the array synthesis for 5 G massive MIMO systems in the line-of-sight working condition. The main result of the numerical investigation performed is that non-uniform arrays are the natural choice in this kind of application. In particular, by using non-equispaced arrays, we show that it is possible to achieve a better average condition number of the channel matrix and a significantly higher spectral efficiency. Furthermore, we verify that increasing the array size is beneficial also for circular arrays, and we provide some useful rules-of-thumb for antenna array design for massive MIMO applications. These results are in contrast to the widely-accepted idea in the 5 G massive MIMO literature, in which the half-wavelength linear uniform array is universally adopted.}


\begin{document}
\section{Introduction}
Massive MIMO is going to become one of the key technologies of the incoming next generation 5 G wireless systems \cite{hoydis2011massive,huh2012achieving,rusek2013scaling,hoydis2013making,nam2013full,hoydis2013massive,lu2014overview,boccardi2014five,jungnickel2014role,larsson2014massive,bjornson2015optimal,bjornson2016massive}. In particular, the use of frequencies above 6 GHz seems to be particularly appealing, because of two main reasons: first of all, the availability of wide contiguous frequency bands to exploit for communications; second, the relative compactness of the resulting arrays \cite{swindlehurst2014millimeter,lee2016randomly,puglielli2016design,bjornson2016massive}. 

If we briefly focus on this second aspect, we would observe that the use of a frequency of the order of 60 GHz would mean a wavelength $\lambda$ of 5 mm; thus, a 100$\lambda$ array would be just half a meter in~size. 

The use of mm-length frequencies is not, unfortunately, without drawbacks. The well-known Friis formula for wireless connections shows a power attenuation that increases with the square of the wavelength, which significantly reduces the area that can be covered by a transmitting antenna. Furthermore, the propagation of the waves at mm-wave frequencies is much different from the propagation at lower frequencies, resembling more a quasi-optical connection between the terminals: the rich scattering condition is really difficult to realize, and the Line-of-Sight (LoS) component becomes dominant (if not the only present) \cite{sarkar2003survey,medbo2014channel}.

This fact has posed some questions on the actual possibility to achieve the so-called ``favorable propagation'' in the Multi-User MIMO (MU-MIMO) channel matrix, but both measurements and simulations have confirmed the advantages of massive MIMO also in the extreme LoS condition~\cite{ngo2014aspects,migliore2014some,gao2015massive,lee2015performance,liu2016channel,chandhar2016ergodic,liu2016geometry}. One of the key points to understand why massive MIMO keeps on working with a very good spectral efficiency also in this case is that the massive antenna array, because of its huge size, does not work in the far-field condition, so that the wave fronts are not planar, but spherical: we could be able to distinguish between two terminals positioned at the same angle with respect to the Base Station (BS) by means of their different distances \cite{zhou2015spherical,chen2016exploration,liu2016benefits}.

Among the many aspects of massive MIMO systems, the optimization of the radiation system has been the object of relatively little research, and equispaced half-wavelength arrays are commonly considered for massive MIMO antennas. However, as pointed out above, massive MIMO antennas work in the near-field condition. This characteristic does not seem to have been exploited in the literature. A further point to be better explored is the reduction of the cost and the complexity of the overall antenna, which is critically related to the number of radiating elements.

Accordingly, a fundamental consideration is now in order: is it possible to exploit this near-field behavior with a fixed complexity of the antenna in terms of the number of elements without affecting, or hopefully increasing, the performance? 
This paper will try to answer this question, showing that optimization of the antenna array using non-uniform arrays in the case of linear antennas and properly equispaced arrays in circular antennas can allow a significant improvement of the communication system performance.

\section{Antennas and Propagation Model}

In this paper, we will use a simplified model for antennas and propagation; in particular, we will consider isotropic radiating elements; we will neglect the effect of mutual coupling and polarization, and we will consider a simple 2D propagation model, i.e., BS antennas and user terminals will lie on the same plane. Furthermore, we will consider a fixed frequency $f$ and the corresponding wavelength~$\lambda$. 

These assumptions have a two-fold reason; first, the model we will discuss, and the achieved results, could be easily reproduced by any interested reader; second, the simplified model will allow us to remove the aspects that play a minor role in the definition of the channel matrix (polarization and mutual coupling), focusing only on the aspects, like the antenna positioning, that play the main role in the overall system performance.

\begin{figure} [H]
\centering
\includegraphics[width=5in]{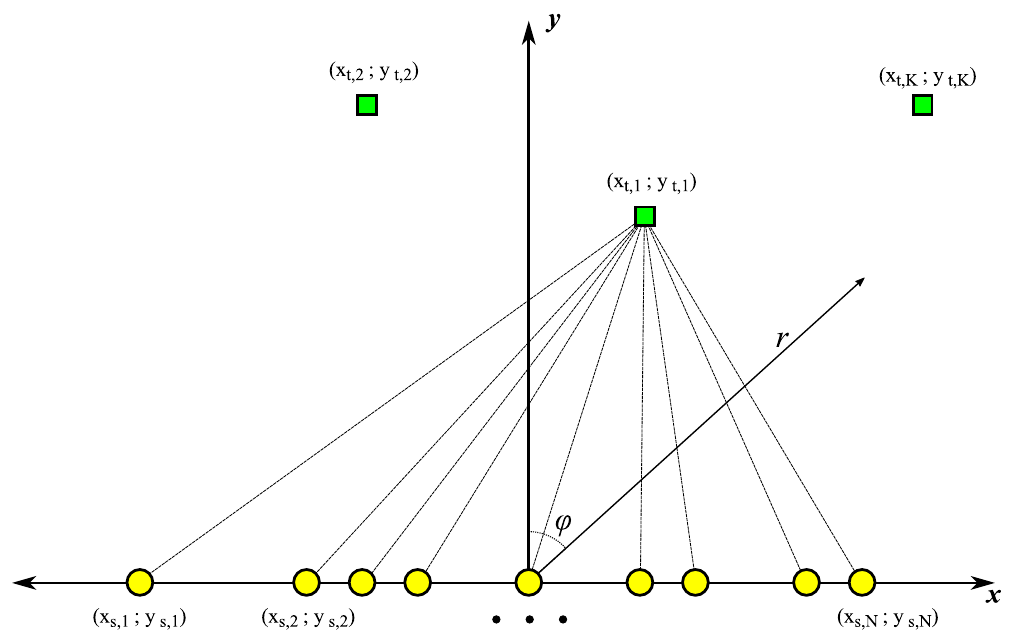}
\caption{Massive MIMO system scenario. Circles: base station antennas; squares: terminals.}
\label{fig:scenario}
\end{figure}

In particular, we will consider a linear antenna array of $N$ identical elements, with the antennas aligned along the $x$ axis, whose abscissas are stored in the vectors $\mathbf x_s=[x_{s,1};\cdots;x_{s,N}]$. The position of the $K$ terminal antennas will be instead described by the vectors $\mathbf x_t=[x_{t,1};\cdots;x_{t,K}]$
and $\mathbf y_t=[y_{t,1};\cdots;y_{t,K}]$ (see Figure \ref{fig:scenario}). The transfer function between the $n$-th antenna of the BS and the $k$-th terminal will be achieved as:
\begin{equation}
h_{n,k}=\gamma \frac{e^{-j \beta R_{n,k}}}{R_{n,k}}
\label{eq:hnk}
\end{equation}
where $\gamma$ is an inessential multiplicative constant, $\beta=2\pi/\lambda$ is the free space wavenumber and: 
\begin{equation}
R_{n,k}=\sqrt{(x_{s,n}-x_{t,k})^2+y_{t,k}^2}
\label{eq:Rnk}
\end{equation}

The MU-MIMO channel matrix can be thus{ built as:}
\begin{equation}
\mathbf{H}=~[ \mathbf{h}_1 \cdots \mathbf{h}_K ]
\label{eq:Hmat}
\end{equation}
where the column vector $\mathbf{h}_k$ {represents} the channel response of the $k$-th terminal. 

This channel matrix can be properly normalized, according to the guidelines in \cite{gao2015massive} before performing its singular value decomposition or calculating the spectral efficiency (thus making the constant $\gamma$ inessential). It has also to be underlined that the use of Equation (\ref{eq:hnk}) allows us to take into account the different power received by the antennas at the base station because of their different distance from the terminals, as well as the spherical shape of the wavefronts.

\section{User Correlation Analysis}
From this section on, we will choose $f=60$ GHz as the working frequency and a base station array of 200 elements. The choice of considering a fixed number of antennas at the BS has been made just to compare different antenna geometries with the same system complexity.

In order to get better insight into the effect of the antenna shape on the system performance, we will start considering the simple case of only two terminals, and we will calculate the correlation coefficient between the two columns of $\mathbf{H}$, as:
\begin{equation}
\chi_{1,2}=\frac{|\mathbf{h}_{1}\cdot\mathbf{h}_{2}^*|}{||\mathbf{h}_{1}|| ~||\mathbf{h}_{2}|| }
\label{eq:corr_h}.
\end{equation}

The use of this performance metric is due to the fact that it provides a simple way to check the effectiveness of a maximal ratio combining approach at the BS; furthermore, as will be clear in the following, it allows us to provide a simple and effective graphical description of the correlation between user terminal positions. 

\begin{figure} [H]
\centering
\includegraphics[width=4in]{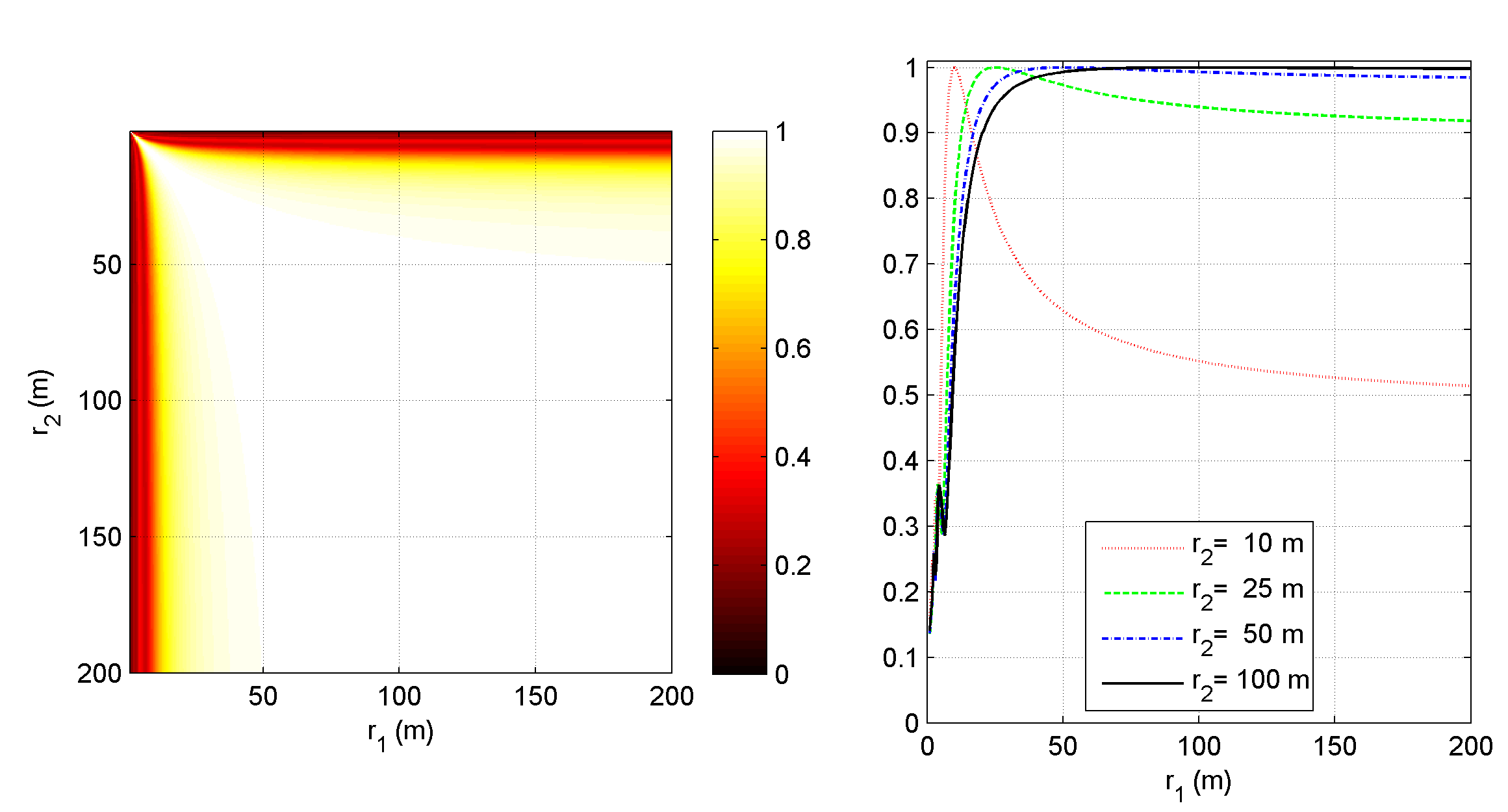}
\caption{Two terminal correlation analysis for the $\lambda/2$ equispaced array. ({Left}) Image map with the correlation level for variable $r_1$ and $r_2$; ({right}): plot of the correlation coefficient for some fixed values of $r_2$.}
\label{fig:EQUI_d05}
\end{figure}

Let us first consider the case of an equispaced array, with the inter-element distance $d=\lambda/2$; in~Figure \ref{fig:EQUI_d05}, it is possible to see the case in which the two terminals are aligned along the $y$ axis, and~have only different distances ($r_1$ and $r_2$) from the BS. In the left subplot, we can see an image map representing the correlation coefficient of Equation (\ref{eq:corr_h}) for variable $r_1$ and $r_2$, while in the right subplot, we can see the plot of the correlation coefficient for some fixed values of $r_2$. This figure confirms that it is possible to distinguish between two users sharing the same angular position with respect to the center of the BS, but this feature can be exploited only when one terminal is very close to the BS with respect to the~other. 

It has to be underlined that this behavior has been obtained when the two terminals lie on the $y$~axis; for other directions, it is possible to obtain different plots, showing a slightly increased correlation, not reported here for the lack of space.

Let us now consider an equispaced array with a larger inter-element distance, say $d=2\lambda$. From~an antenna point of view, increasing the inter element distance, without changing the number of elements, increases the far-field distance of the array, which is conventionally calculated as:
\begin{equation}
r_{FAR}=\frac{2 [(N-1) d]^2}{\lambda}.
\end{equation}

\begin{figure} [H]
\centering
\includegraphics[width=4in]{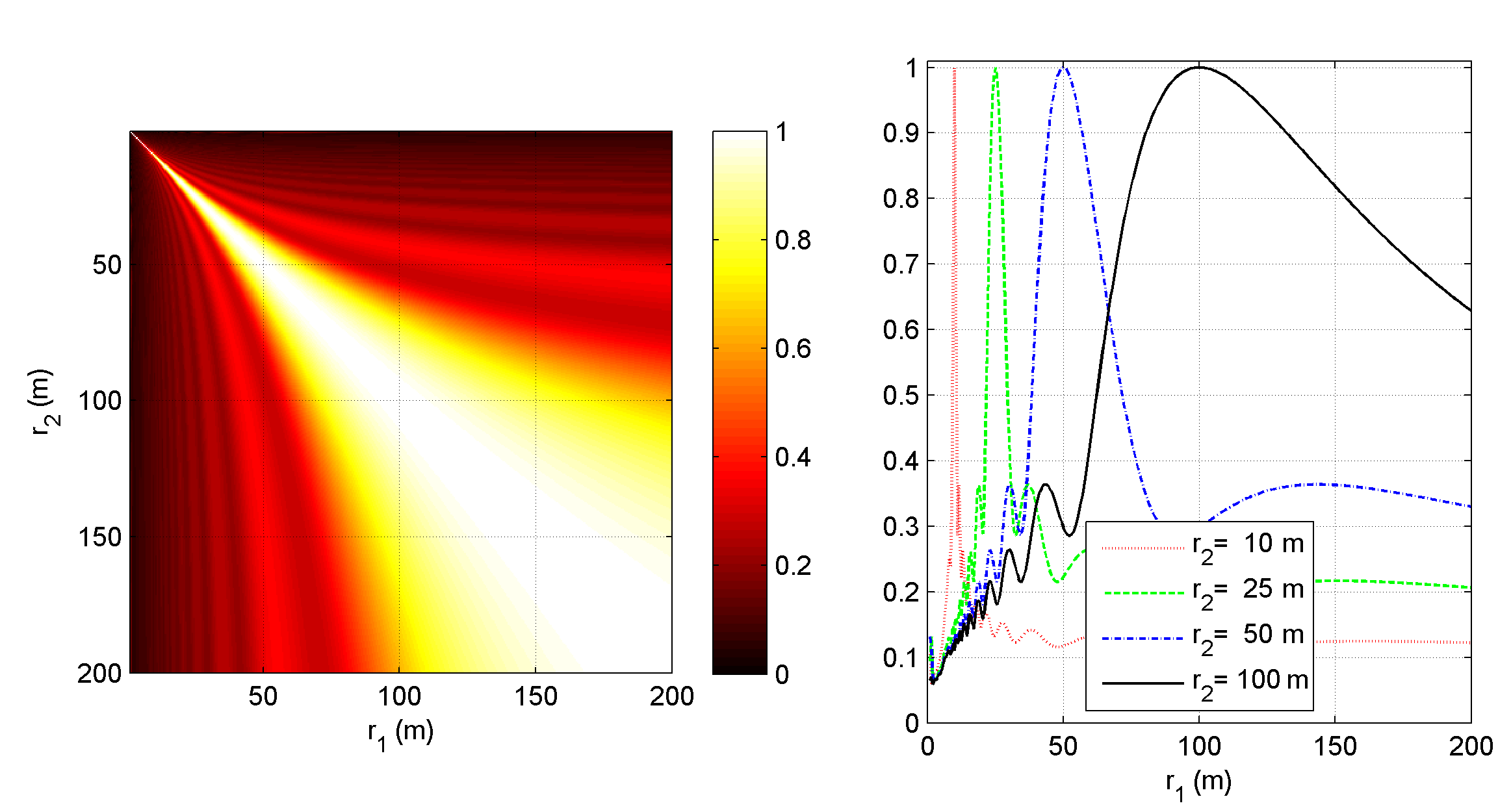}
\caption{Two terminal correlation analysis for the $2\lambda$ equispaced array. ({Left}) Image map with the correlation level for variable $r_1$ and $r_2$; ({right}): plot of the correlation coefficient for some fixed values of $r_2$.}
\label{fig:EQUI_d2}
\end{figure}

This means that an increase of the inter-element distance of a four factor increases the far field distance of a 16 factor. In Figure \ref{fig:EQUI_d2}, we can see the image map and the plot of the correlation for the larger array. It is clear that we are now allowed to easily distinguish two users that share the same angular position with respect to the BS, much more than in the previous case.
\vspace {-12pt}
\vspace {-18pt}

\begin{figure} [H]
\centering
\includegraphics[width=4in]{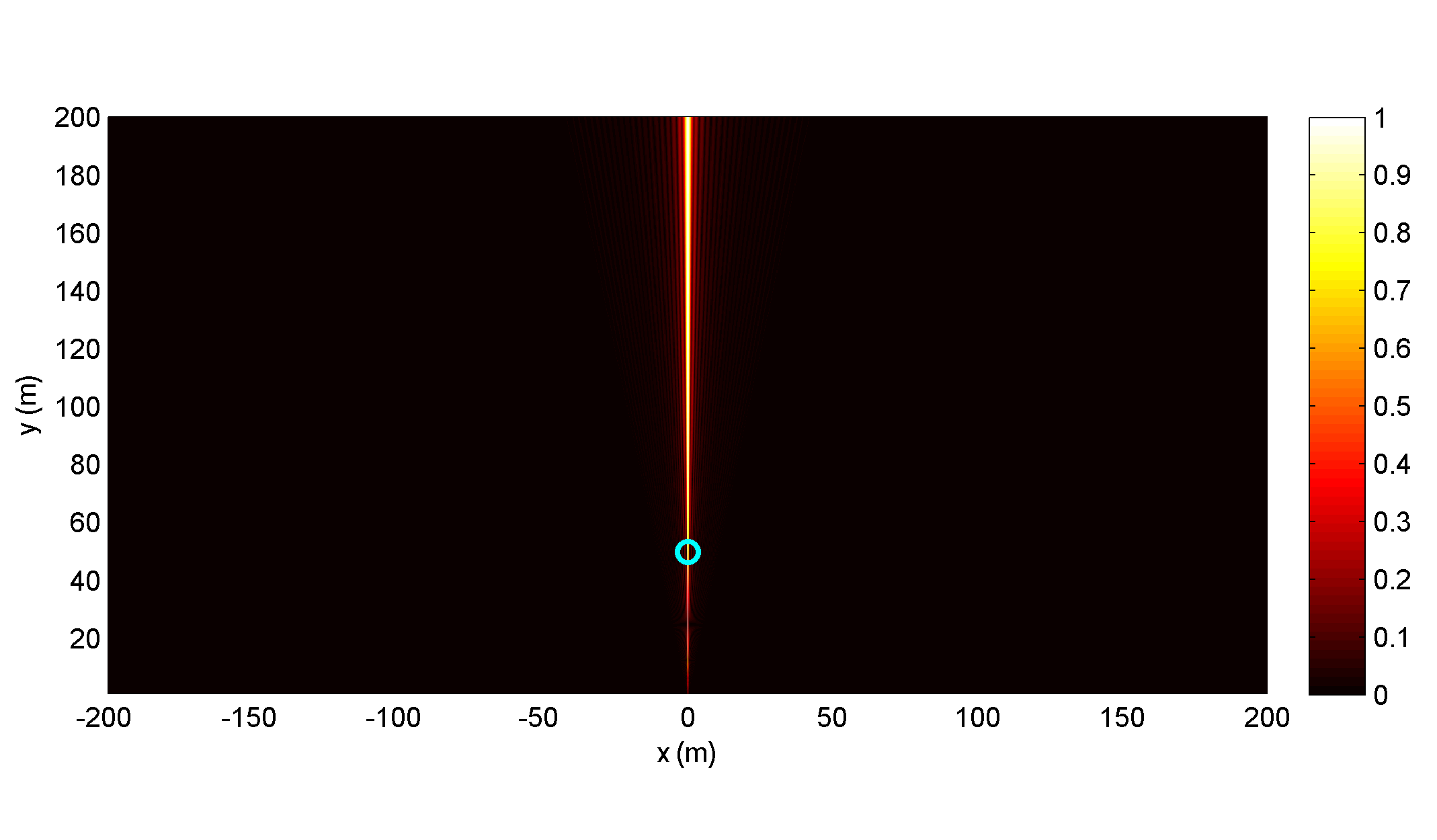}
\caption{Image map of the correlation coefficient when the first terminal is fixed in the position marked with the circle and the second user is moved on the plane for the $\lambda/2$ equispaced array.}
\label{fig:PIANO_d05}
\end{figure}

\begin{figure} [H]
\centering
\includegraphics[width=4in]{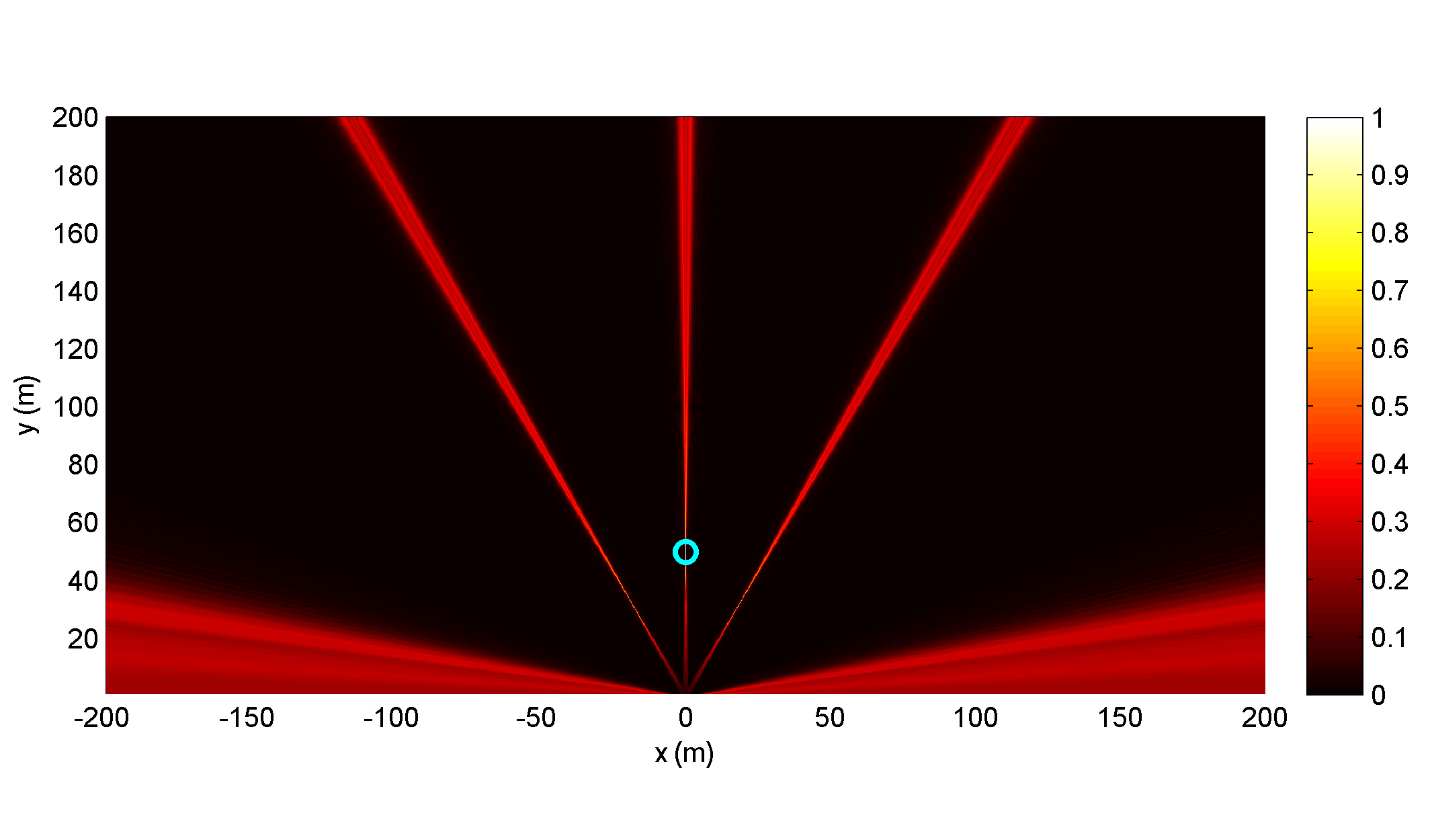}
\caption{Image map of the correlation coefficient when the first terminal is fixed in the position marked with the circle and the second user is moved on the plane for the $2\lambda$ equispaced array.}
\label{fig:PIANO_d2}
\end{figure}

The previous result has shown that the increase of the inter-element distance seems to have a beneficial effect on the capability of decorrelating users. To confirm this feature, let us see what happens if we consider the first terminal in a fixed position with respect to the array, and we move the second terminal on the $xy$ plane. In Figure \ref{fig:PIANO_d05}, we can see the case of a $\lambda/2$ equispaced array: apart~from an angular sector around the broadside direction of the array, the correlation is very low, thus confirming the very good results, in terms of favorable propagation, found by other researchers. If we instead employ the $2\lambda$ equispaced array, we obtain the result depicted in Figure \ref{fig:PIANO_d2}: apart from the broadside direction, which shows a slightly lower correlation level with respect to the $\lambda/2$ case, there~are other directions in space that present a non-negligible correlation. These regions are related to the unavoidable grating lobes, appearing when we consider equispaced elements with an inter-element distance greater than half-wavelength. Clearly, the presence of grating lobes causes ambiguities and must be avoided. As an example, let us imagine a communication system with just two user terminals: when user Terminal A is in the angular position of a grating lobe relative to user Terminal B, it could be very difficult to distinguish them.

\unskip

We would like to specify that in an array not working in the far field region, we could not rigorously talk of grating lobes, but for the sake of simplicity, we will keep on using this definition for the rest of the manuscript.

According to the correlation analysis, it seems that the possible advantage of distinguish aligned user terminals could be negated by the effect of grating lobes. These results suggest that a non-equispaced array, not affected by grating lobes, could get the advantages of a larger size of the array, without the drawbacks of the grating lobes, but to confirm this hypothesis, we are first required to analyze what happens with multiple terminals.

\section{Multi User Channel Matrix}

Let us now consider a number $K$ of terminals, uniformly randomly displaced in an angular sector $\varphi\in[-\varphi_s,\varphi_s]$, with a minimum distance $r_{min}$ and a maximum distance $r_{max}$ from the center of the BS array. For all of the simulations that will be discussed in this section, we will use $\varphi_s=\pi/3$, $r_{min}=5$~m and $r_{max}=200$ m; the choice of these parameters has been done in order to simulate a sectorized coverage of a large planar area, like a plaza or a park. It has to be underlined that differently from~\cite{masouros2015space}, we will consider a fixed number of antennas at the BS, in order to compare systems with the same complexity (and cost).

In order to better emphasize the possibility to discriminate different users according to their distance, we will employ ``Normalization 1'' from \cite{gao2015massive}. In this way, all of the columns of the channel matrix $\mathbf{H}$ will have the same norm, thus removing the unbalance of channel attenuation between users due to their different relative distance from the BS. $M=1000$ different random scenarios have been considered in the numerical simulations.

\begin{figure} [H]
\centering
\includegraphics[width=4in]{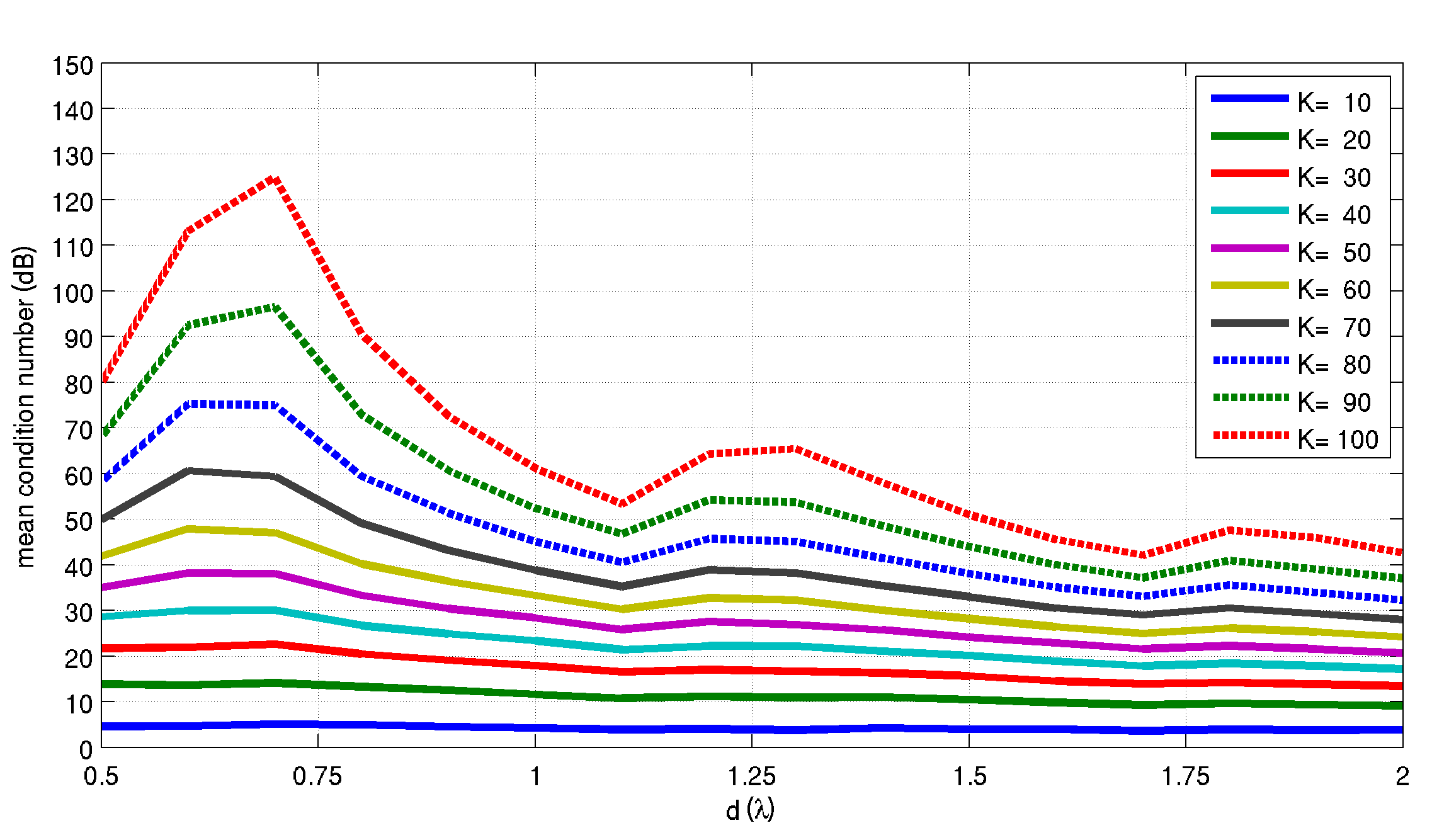}
\caption{Mean condition number of the channel matrix as a function of the inter-element distance $d$ and the number of users $K$.}
\label{fig:MeanCond_x1}
\end{figure}

As a first test, in Figure \ref{fig:MeanCond_x1}, we show as a function of the inter-element distance $d$ and the number of users $K$ the average condition number of the channel matrix, defined as:
\begin{equation}
\rho_{dB}=20*log_{10}\left(\frac{\sigma_1}{\sigma_K}\right)
\end{equation}
where $\sigma_1$ and $\sigma_K$ are the greatest and smallest singular values of the channel matrix. The choice of the condition number is due to the fact that it provides a direct indication of the orthogonality of the columns of the channel matrix, and it is related to the stability of its inversion \cite{golub2012matrix}, needed in many multi-user MIMO transmission schemes \cite{tsoulos2006mimo}.

The main advantage in using the condition number is that the presented results are independent of the actual SNR at the receiver; it is indeed true that the condition number provides only a rough indication of the channel quality, since it gives only the spread of the singular values and not their actual distribution, but its value is strongly related with the performances of MIMO
precoders and detectors \cite{artes2003efficient,maurer2007low,mohammed2011mimo}.

We see that the average condition number is an increasing function of the number of users, but it is interesting to note that (besides some oscillations), it is also a decreasing function of the inter-element distance $d$ at the base station, confirming the advantages of a larger array at the BS foreseen in the previous paragraph. 

A brief discussion of the oscillations is now needed. Taking as a reference the $K=100$ curve of Figure \ref{fig:MeanCond_x1}, we can see that the condition number increases up to about $d=0.7\lambda$, then it decreases up to $d=1.1\lambda$, then it increase again up to $d=1.3\lambda$, and so on. As we have seen in Figure \ref{fig:PIANO_d2}, the presence of grating lobes implies that a number of angular regions of the scenario should be avoided for an effective communication. The number of grating lobes that affect the results is indeed an increasing discrete function of $d$, while the reduction of the beamwidth and the correlation among users sharing the same angle $\varphi$ with respect to the BS are instead a continuous decreasing function of $d$. These two effects together contribute to the aforementioned oscillating/decreasing behavior seen in Figure \ref{fig:MeanCond_x1}.

It is important to recall that a very low condition number is necessary to apply the Maximal Ratio Combining (MRC) algorithm at the BS, and a low condition number is also necessary to correctly calculate the matrix inversion needed to efficiently use Zero Forcing (ZF) at the BS \cite{yang2013performance,bjornson2015optimal}.

\begin{figure} [H]
\centering
\includegraphics[width=5in]{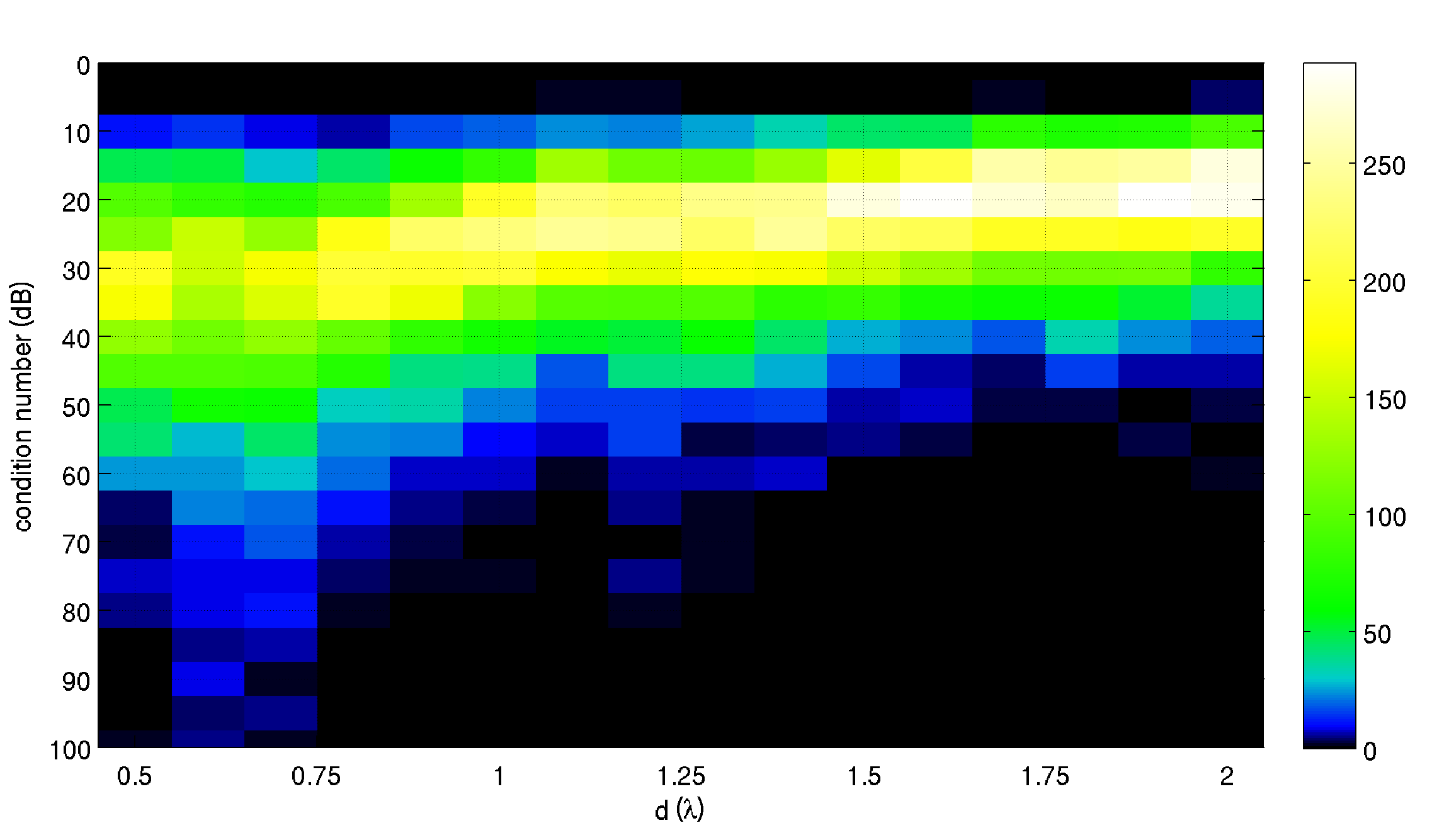}
\caption{Distribution of the condition number as a function of the distance $d$ for $K=50$. Each column of the matrix in the image map is a histogram of the condition number.}
\label{fig:HistCond50_x1}
\end{figure}

To get better insight into this result, let us focus on the case of $K=50$; in Figure \ref{fig:HistCond50_x1}, we show the distribution of the condition number as a function of the distance $d$ (each column of the matrix in the image map is a histogram of the condition number). The increase of $d$ from $\lambda/2$ to $2\lambda$ reduces the average value of the condition number of about 15 dB, and also, its spreading decreases.

\begin{figure} [H]
\centering
\includegraphics[width=5in]{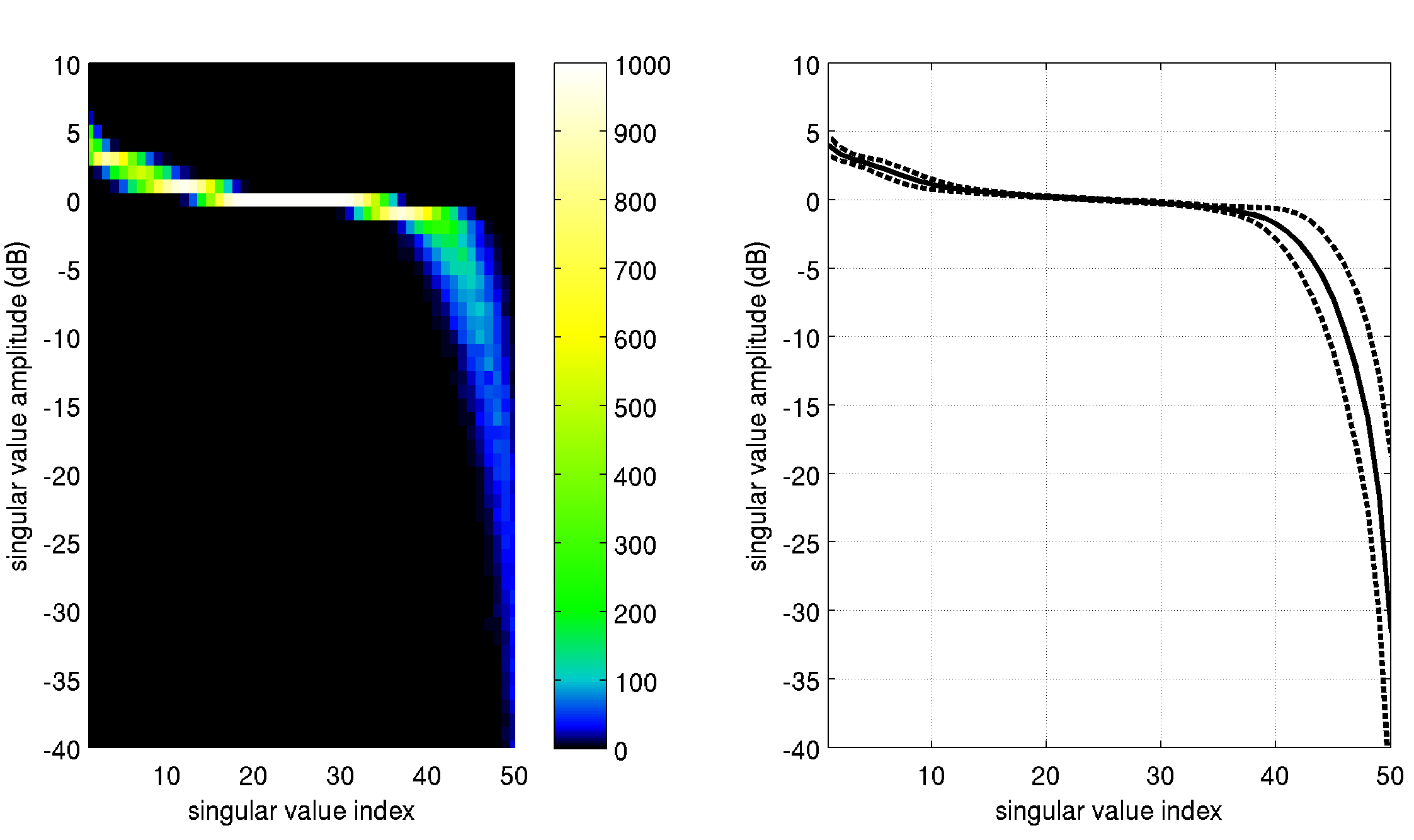}
\caption{Singular values analysis for $d=\lambda/2$ and $K=50$. ({Left}) Image map of the distribution of each one of the 50 singular values; ({right}) solid line, average singular values; dashed lines, average singular values $\pm$ the standard deviation of the distribution.}
\label{fig:PROVV_HistVS_d05_x1}
\end{figure}

\begin{figure} [H]
\centering
\includegraphics[width=4.5in]{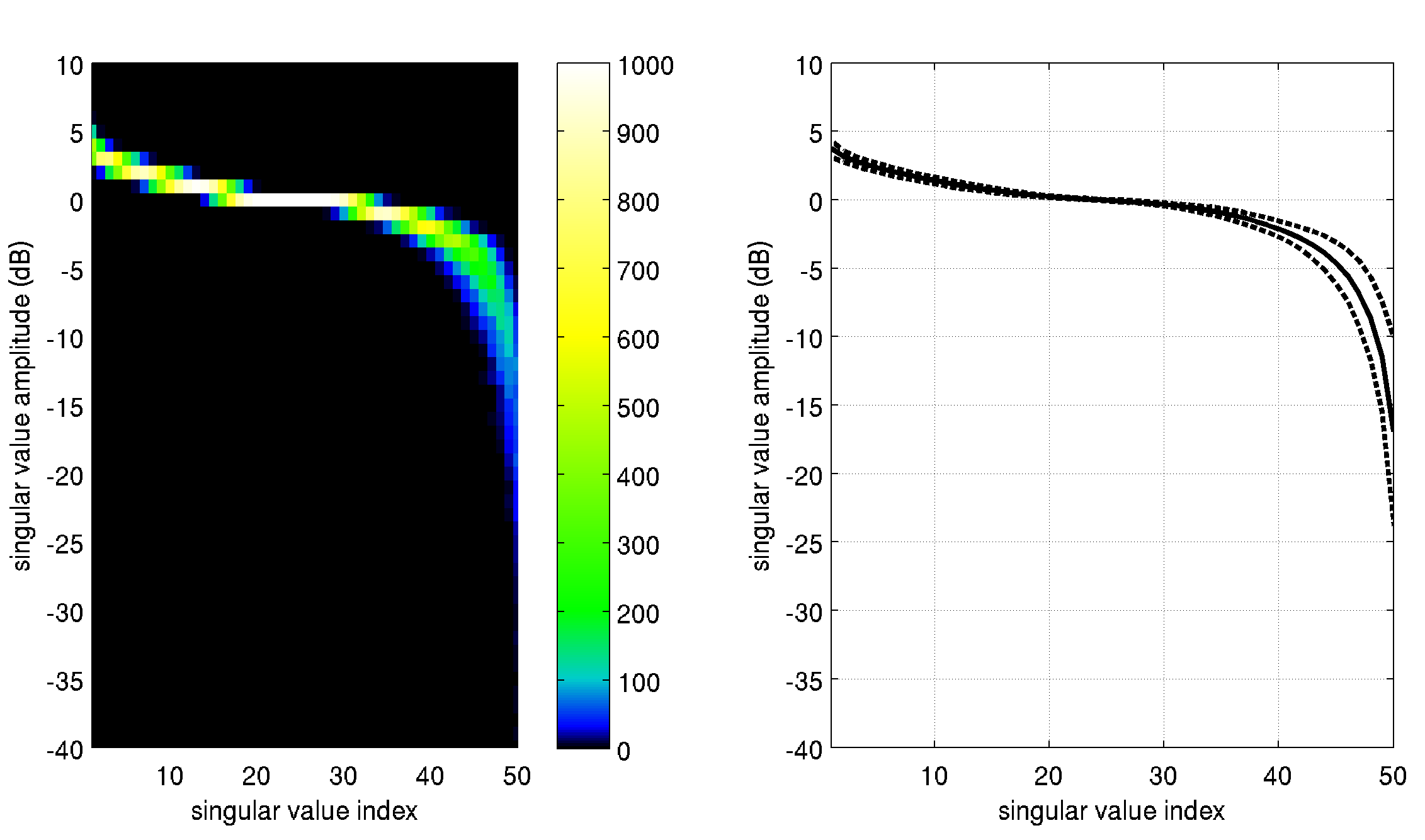}
\caption{Singular values analysis for $d=2\lambda$ and $K=50$. ({Left}) Image map of the distribution of each one of the 50 singular values. ({Right}) solid line, average singular values; dashed lines, average singular values $\pm$ the standard deviation of the distribution.}
\label{fig:PROVV_HistVS_d2_x1}
\end{figure}

With reference to the two extreme cases of $d=\lambda/2$ and $d=2\lambda$, let us now analyze the distribution of the singular values for $K=50$.
In Figures \ref{fig:PROVV_HistVS_d05_x1} and \ref{fig:PROVV_HistVS_d2_x1}, we have the image map of the distribution of the singular values, as well as the plot of their average values. The larger spacing improves in particular the lower singular values, leading to a much more stable inversion. 

It has to be underlined that the results achieved in this paragraph hold also when we do not consider the normalization. Since we would just see a slightly larger spreading of the singular values for all of the considered cases, we do not report these results because of the limited space of the paper.

\section{Exploiting Sparse Arrays}
In the previous section, we have shown that increasing the size of the array is beneficial to the overall MU-MIMO channel matrix condition number: even if we have to deal with the effect of grating lobes, the possibility to discriminate sources that share a similar angular direction, as well as the reduced width of the main beam lead to a better conditioned matrix.

It is trivial to recognize that, if we could get rid of the grating lobes, we should expect a further improvement of the performances. To this aim, we would have two different possibilities: in the first case, we could employ a sub-array architecture \cite{mailloux2005phased}, but we should also deal with the reduced scanning capability of our array (and hence, the dimension of the sector we could cover); the second possibility would be to employ a non-equispaced, sparse array.

The synthesis of sparse arrays is one of the most studied antenna topics in the last 50 years, and~a very large number of methods and algorithms has been developed for linear and planar arrays~\cite{donelli2004linear,massa2004planar,liu2008reducing,oliveri2009linear,poli2013reconfigurable,oliveri2014sparsening,bucci2015synthesis,d2016maximally,pinchera2016compressive,rocca2016unconventional,pinchera2016comparison,bucci2017interleaved}. Unfortunately, none of these algorithms can be directly applied to an MU-MIMO system: our~aim is to obtain the antenna positions, within a given range, that allow the best performances of a multi-user scenario, and this problem cannot be directly translated into a classical antenna pattern synthesis~problem.

To solve this issue, we propose a novel paradigm for antenna array synthesis, inspired by the results presented in the previous paragraph: we will be seeking for the antenna positions providing the best performances on a relatively large set of user terminal scenarios. We could, for example, optimize the antenna array to provide the lowest average condition number of the channel matrix, or~we could look for the antenna configuration that guarantees the lowest 0.99-quantile of the condition number.

Once we have defined our optimization objective, we could use an evolutionary search method to achieve the sought antenna positions. Actually, the direct application of this approach could be very ineffective, since the number of unknowns we have to deal with is of the order of the number of antennas, and evolutionary algorithms are known to require a computational time that increases exponentially with the number of unknowns \cite{nemirovskii1983problem}. 

A much more convenient, yet sub-optimal, choice is to define a proper parametrization of the antenna positions, thus reducing the search space of the evolutionary algorithm to a small number of parameters and leading to a reasonable computational time.

A simple and effective parametrization of the antenna positions can be found using Tchebyshev polynomials. If we are looking for a symmetric array, the position of the $n$-th antenna ${x}_{s,n}$ could be found has:
\begin{equation}
{x}_{s,n}=L \left( \left(1-\sum_{p=1}^{P}\alpha_p\right) u_n +\sum_{p=1}^P \alpha_p T_{2p+1}(u_n) \right)
\label{eq:TchSparse}
\end{equation}
with:
\begin{equation}
L=\frac{d_0}{2(N-1)}
\end{equation}
where $T_p(\cdot)$ is the Tchebyshev polynomial of order $p$, $[\alpha_1,\cdots,\alpha_P]$ is the set of $P$ parameters defining the position of the antenna elements, $d_0$ is the desired average inter-element distance, $N$ is the overall number of antennas{ and} $[u_1;\cdots;u_N]$ {is an} equispaced ordered sampling of the interval $[-1,1]$. Not all of the {combinations of} $[\alpha_1;\cdots;\alpha_P]$ {would} lead to a monotonically-increasing vector of positions $\mathbf x_s$ in the range $[-L,L]$; for instance, it is easy to demonstrate that if $P=1$, any value of $\alpha_1$ in the range: 
\begin{equation}
-\frac{1}{8}<\alpha_1<\frac{1}{4}
\label{eq:alpharange}
\end{equation}
would lead to a set of positions in the range $[-L,L]$ satisfying the monotonicity constraint, but for higher values of $P$, it would be non-trivial to achieve a formula like Equation (\ref{eq:alpharange}) in closed form.

In the remaining part of the paper, we will focus on the relatively easy case of $P=1$; this choice is justified by the results of some numerical tests, which have shown only a marginal improvement when using $P>1$. In particular, values of $\alpha_1<0$ would lead to antenna arrays with a higher density of elements at the extrema, while values of $\alpha_1>0$ would lead to antenna arrays with a higher density of elements at the center.

\begin{figure} [H]
\centering
\includegraphics[width=4in]{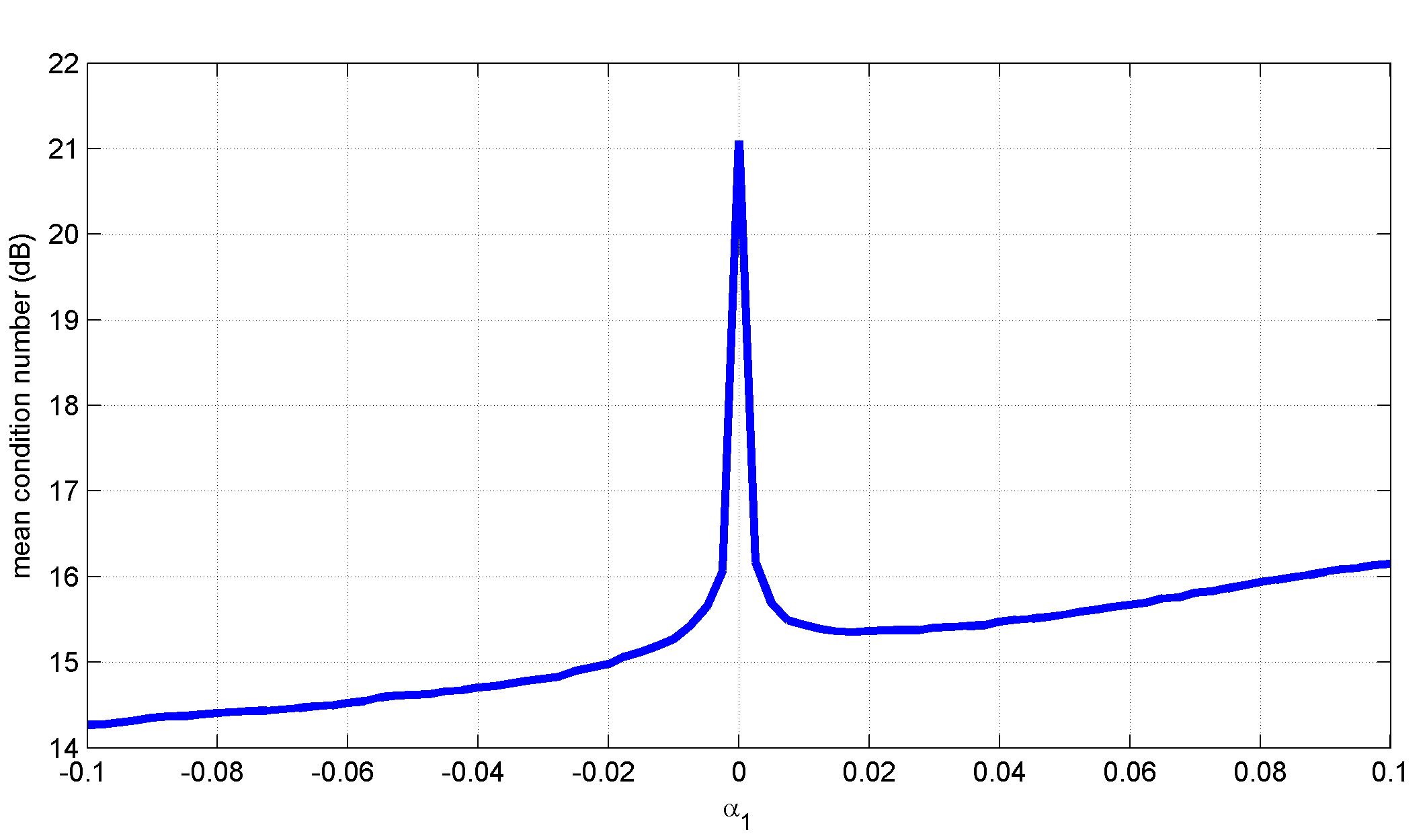}
\caption{Average condition number as a function of $\alpha_1$ for a number of users $K=50$ and $d_0=2\lambda$.}
\label{fig:delta0_parametric_d2lam50}
\end{figure}

Following the guidelines discussed before, we will consider now a parametric analysis of the performances, in terms of average condition number, on a set of 1000 configurations of $K=50$ user terminals, as a function of $\alpha_1$. The result is depicted in Figure \ref{fig:delta0_parametric_d2lam50}: apart from values very close to $\alpha_1=0$, corresponding to the equispaced array, we achieve a significant advantage in the mean condition number for both positive and negative values of $\alpha_1$ (with a slight advantage for the latter). 

To get better insight into this surprising result and to better understand the effect of the non-linear spacing on the array performances, we will consider the same analysis developed in the previous two paragraphs on a test array obtained with $\alpha=-0.03$ with an average inter-element distance $d_0=2\lambda$. The choice of this particular value of $\alpha_1$ is driven by the desire of reducing the spreading of the inter-element distance, in particular in this case, we have a minimum inter-element distance (at the edges) of $1.53\lambda$ and a maximum inter-element distance (at the center) of $2.24\lambda$.

\begin{figure} [H]
\centering
\includegraphics[width=4in]{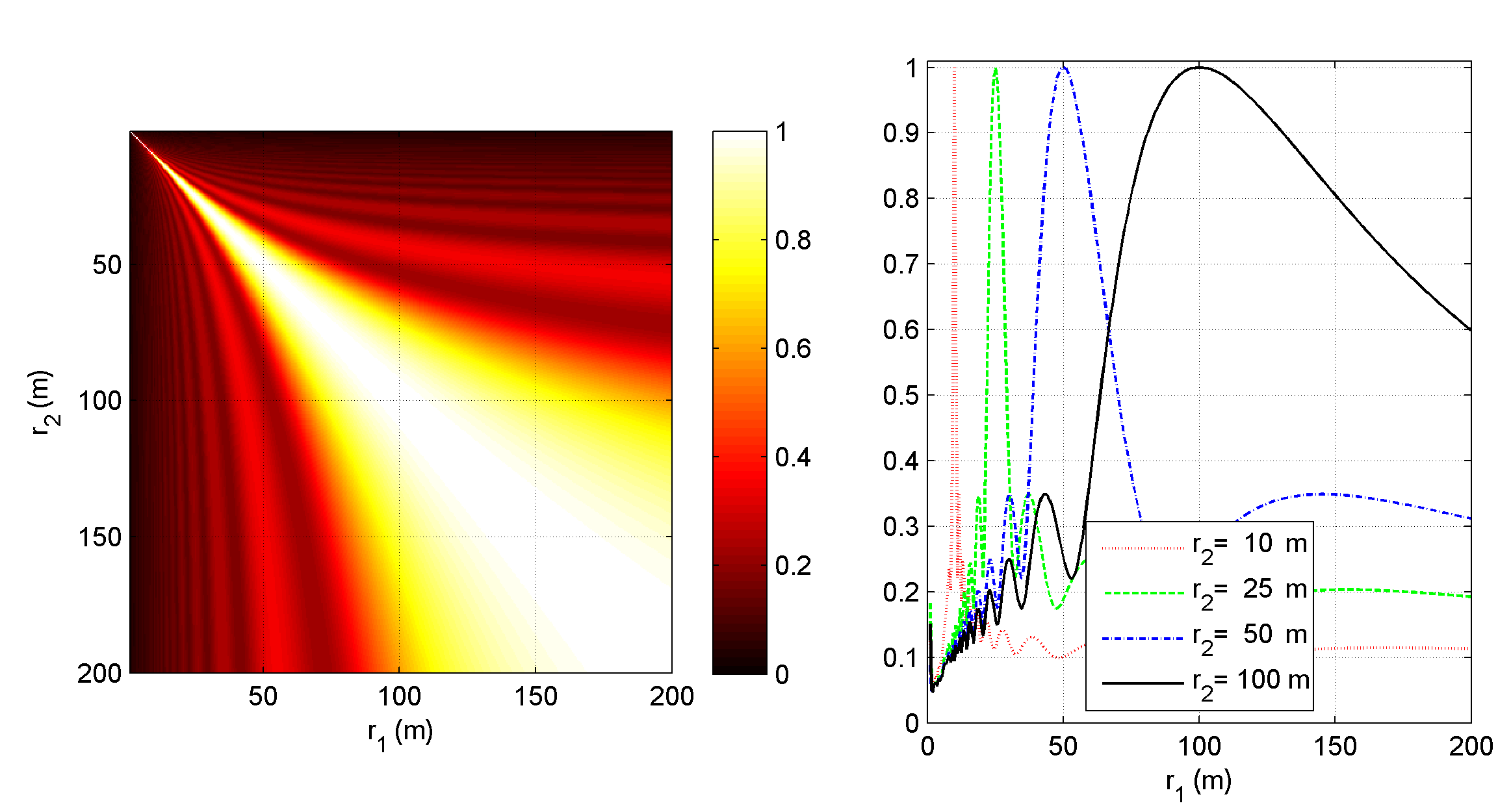}
\caption{Two terminal correlation analysis for the $2\lambda$ non-equispaced array with $\alpha_1=-0.03$. ({Left}) Image map with the correlation level for variable $r_1$ and $r_2$; ({right}) plot of the correlation coefficient for some fixed values of $r_2$.}
\label{fig:SPARSE003_d2}
\end{figure}

\begin{figure} [H]
\centering
\includegraphics[width=4in]{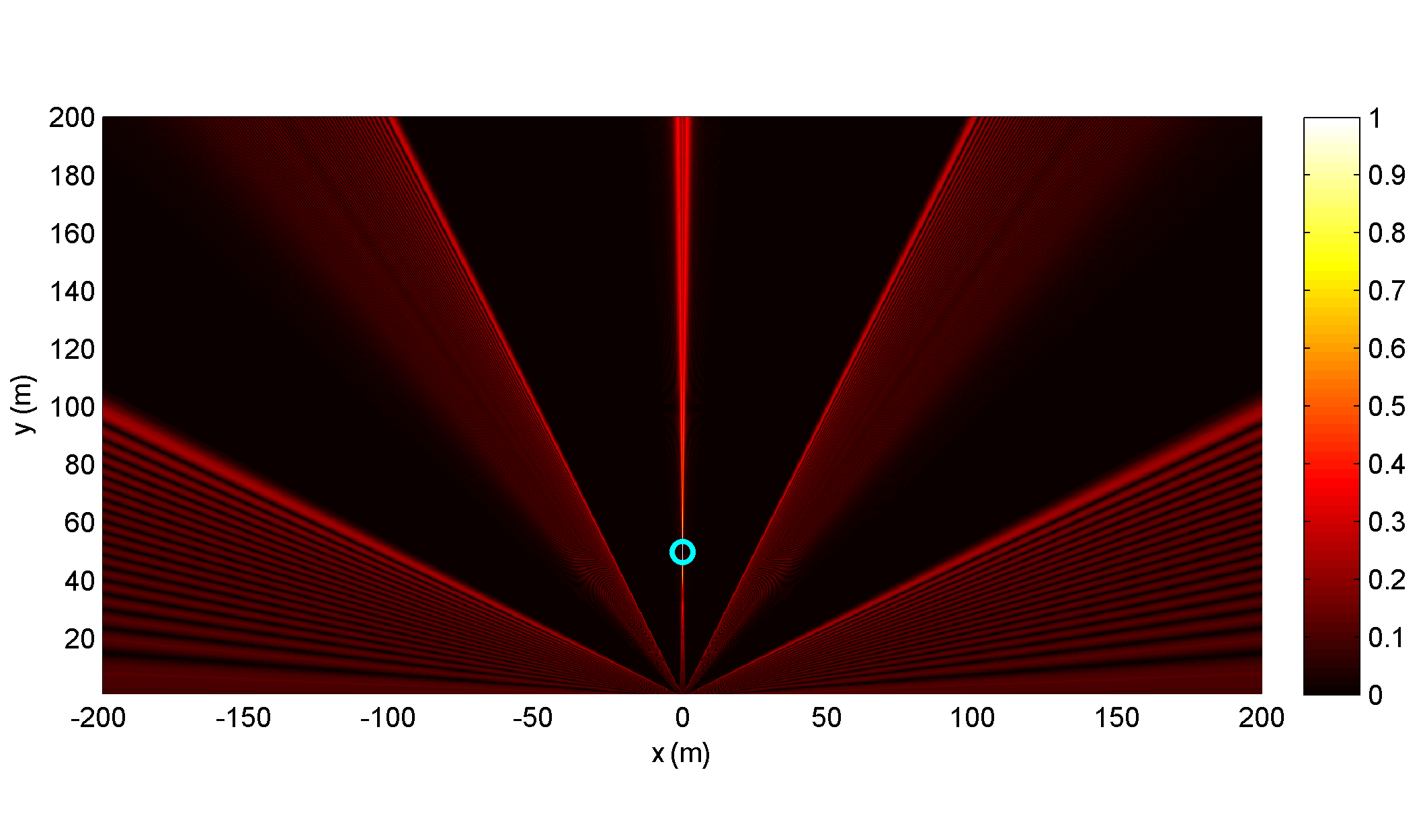}
\caption{Image map of the correlation coefficient when the first terminals are fixed in the position marked with the circle and the second user is moved on the plane for the $2\lambda$ non-equispaced array with $\alpha_1=-0.03$.}
\label{fig:PIANO003_d2}
\end{figure}

In Figure~\ref{fig:SPARSE003_d2}, we can see that the capability of the array to distinguish terminals sharing the same angular position with respect to the BS has been marginally improved with respect to the case of Figure~\ref{fig:EQUI_d2}. The difference with respect to the corresponding equispaced array is instead much clearer from the analysis of Figure~\ref{fig:PIANO003_d2}: now, the effect of the grating lobes is replaced by some sectors of the plane that show a lower (but non-null) correlation coefficient.

\unskip

\begin{figure} [H]
\centering
\includegraphics[width=4in]{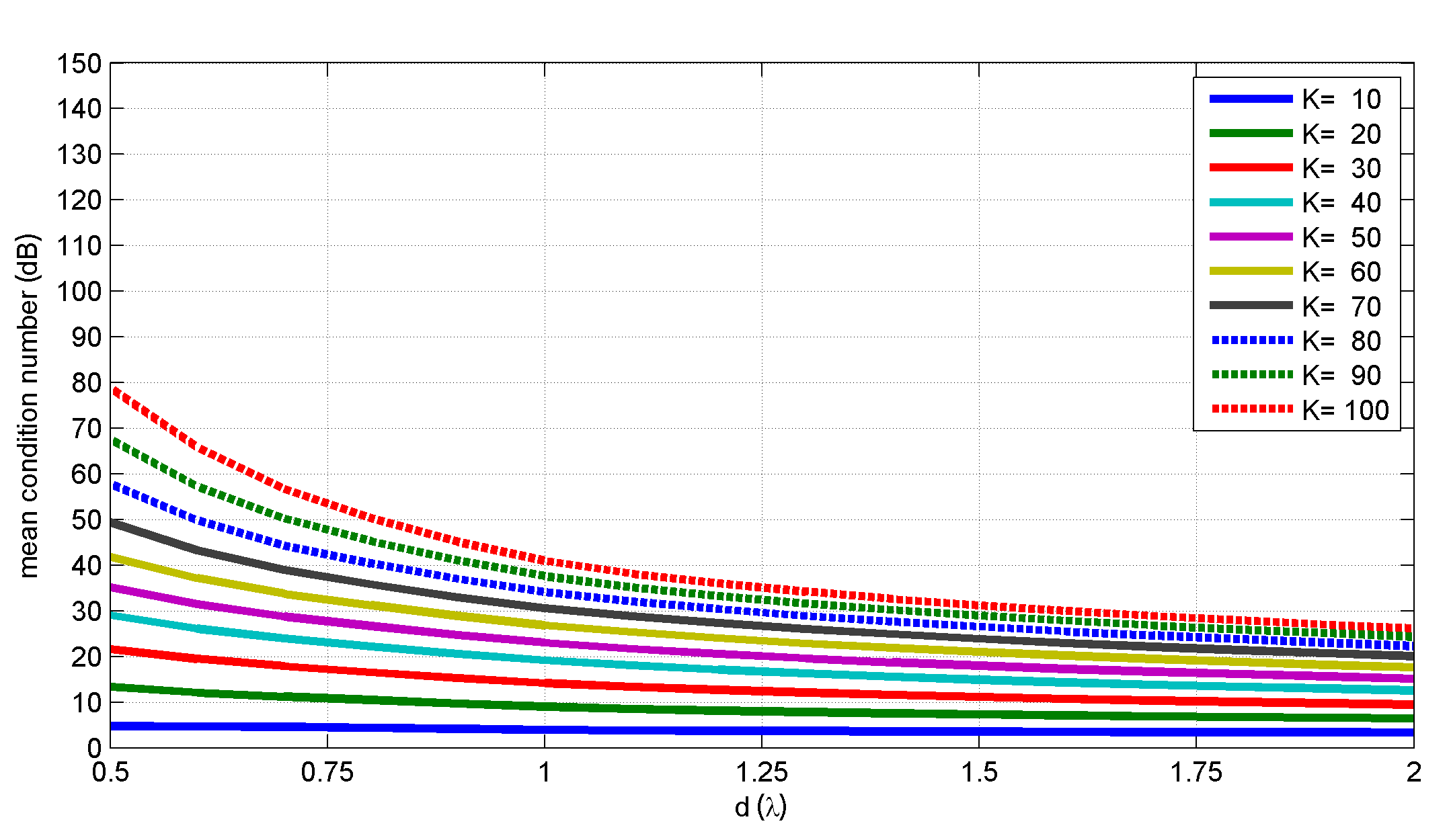}
\caption{Mean condition number of the channel matrix as a function of the average inter-element distance $d$ and the number of users $K$ for a non-equispaced array with $\alpha_1=-0.03$.}
\label{fig:MeanCond_x097}
\end{figure}

In Figure \ref{fig:MeanCond_x097}, we have instead depicted the effect of the non-linear spacing for a variable number of the average inter-element distance $d$ and the number of users $K$: the oscillatory behavior of Figure \ref{fig:MeanCond_x1} has now disappeared, and the condition number is now a monotonic function of the number of terminals and the average inter-element distance. The improvement with respect to the equispaced array is even more significant when the number of user terminals is higher. It is worth underlining that the use of a non-equispaced array is beneficial even when using an inter-element distance only slightly higher than half-wavelength, so to achieve a performance improvement, we are not required to employ large array~sizes.

\begin{figure} [H]
	\centering
	\includegraphics[width=4in]{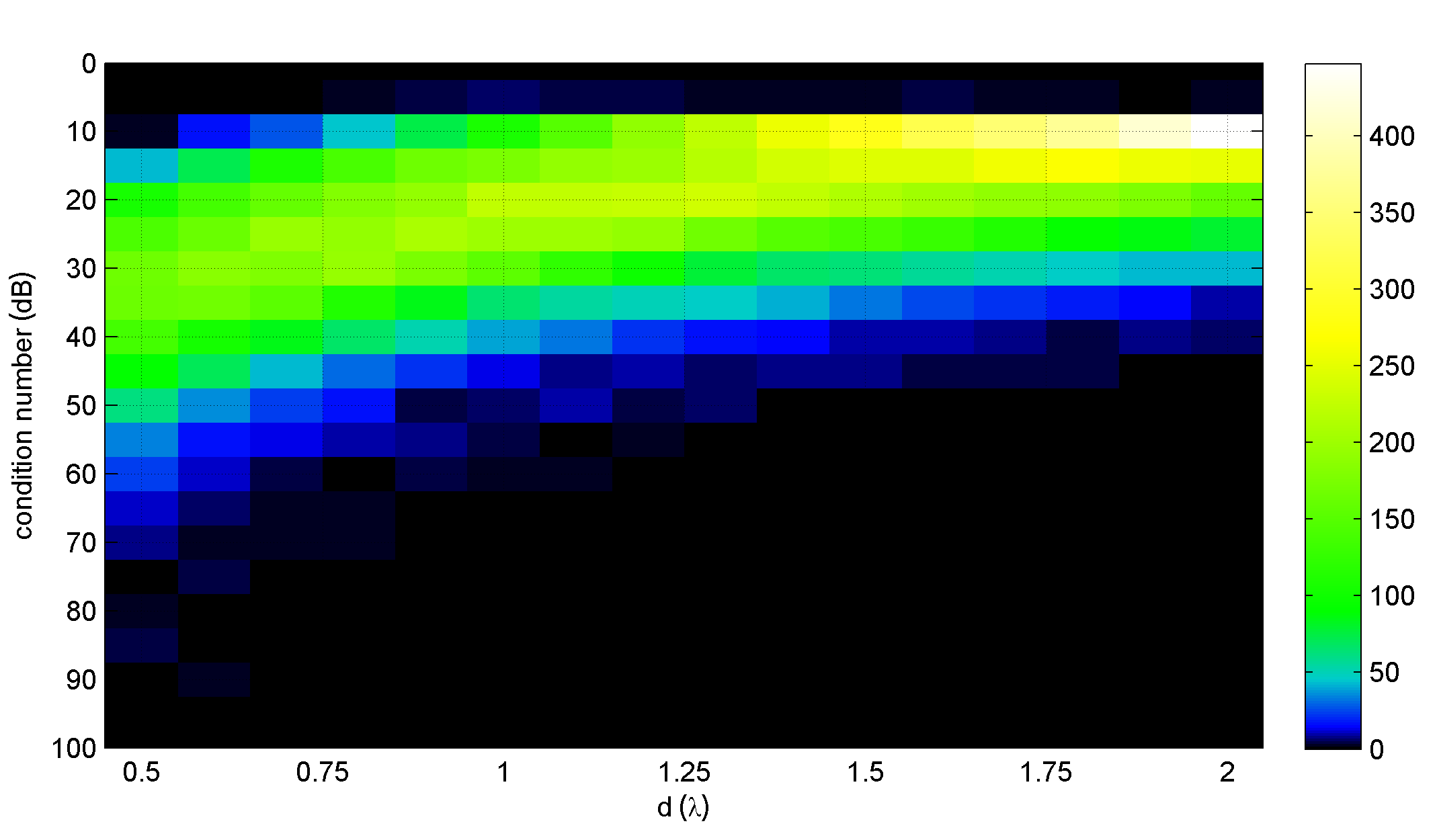}
	\caption{Distribution of the condition number as a function of the average distance $d$ for $K=50$ and the sparse array with $\alpha_1=-0.03$. Each column of the matrix in the image map is a histogram of the condition number.}
	\label{fig:HistCond50_x097}
\end{figure}

\begin{figure} [H]
	\centering
	\includegraphics[width=4in]{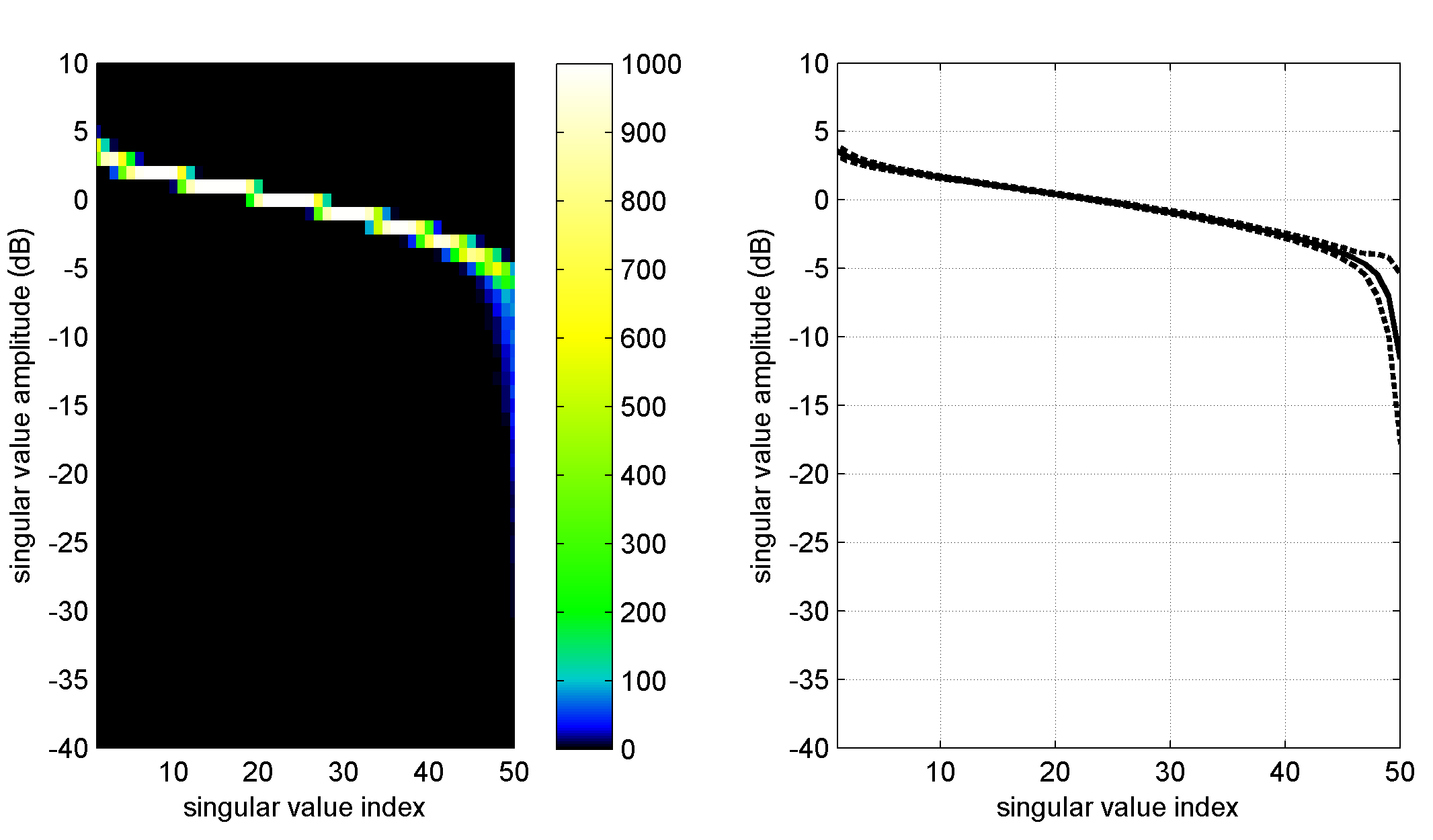}
	\caption{Singular values analysis for $d_0=2\lambda$, $\alpha_1=-0.03$ and $K=50$. ({Left}) Image map of the distribution of each one of the 50 singular values; ({right}) solid line, average singular values; dashed lines, average singular values $\pm$ the standard deviation of the distribution.}
	\label{fig:HistVS_d2_x097}
\end{figure}

It is also interesting to see that the spreading of the condition number is reduced when using a non-equispaced array: the result in Figure \ref{fig:HistCond50_x097}, where the case with $K=50$ is analyzed, shows a conditioning significantly better than the one presented in the equispaced case of Figure \ref{fig:HistCond50_x1}, for any average inter-element spacing. Finally, also the analysis of the distribution of the singular values for the $K=50$ case confirms the advantages of the non-equispaced array (see Figure \ref{fig:HistVS_d2_x097}).

\unskip
\unskip

It is important to underline that, whatever the optimization approach used to define the position of the elements in the sparse array, those positions are fixed: their choice has to be done only once, when designing the array. 

It is also worth underlining that in the simulations shown, no smart terminal selection techniques~\cite{ngo2014aspects} have been employed. This means that we can achieve an even better conditioned channel matrix by the application of these approaches.

\section{Circular Arrays}
In the previous sections, we have considered linear BS arrays for a sectoral coverage; if the terminals are displaced in a circular area, it could be convenient to use a circular array. The increase of the antenna dimension, without changing the number of active elements, can be very beneficial also in this case.

A circular array does not show grating lobes, so it will not be necessary to consider non-equispaced positioning of the array elements on the circle, but the increase in size would allow a better discrimination of the users, because of the increased far field distance of the array.

To better compare the results of the present section to the results of the previous ones, we will always consider $N=200$ radiating elements, displaced on a circle of radius $R=Nd/(2\pi)$ where $d$ is the (approximate) inter-element distance.

For the sake of compactness of the paper, we will skip the correlation analysis, and we will only focus on the condition number analysis of the MU-MIMO channel matrix. In particular, we will consider a set of 1000 different propagation scenarios, where a variable number $K$ of randomly-positioned user terminals is displaced between a distance $r_{min}=5$ m and $r_{max}=200$ m from the center of the BS antenna.

\begin{figure} [H]
\centering
\includegraphics[width=4in]{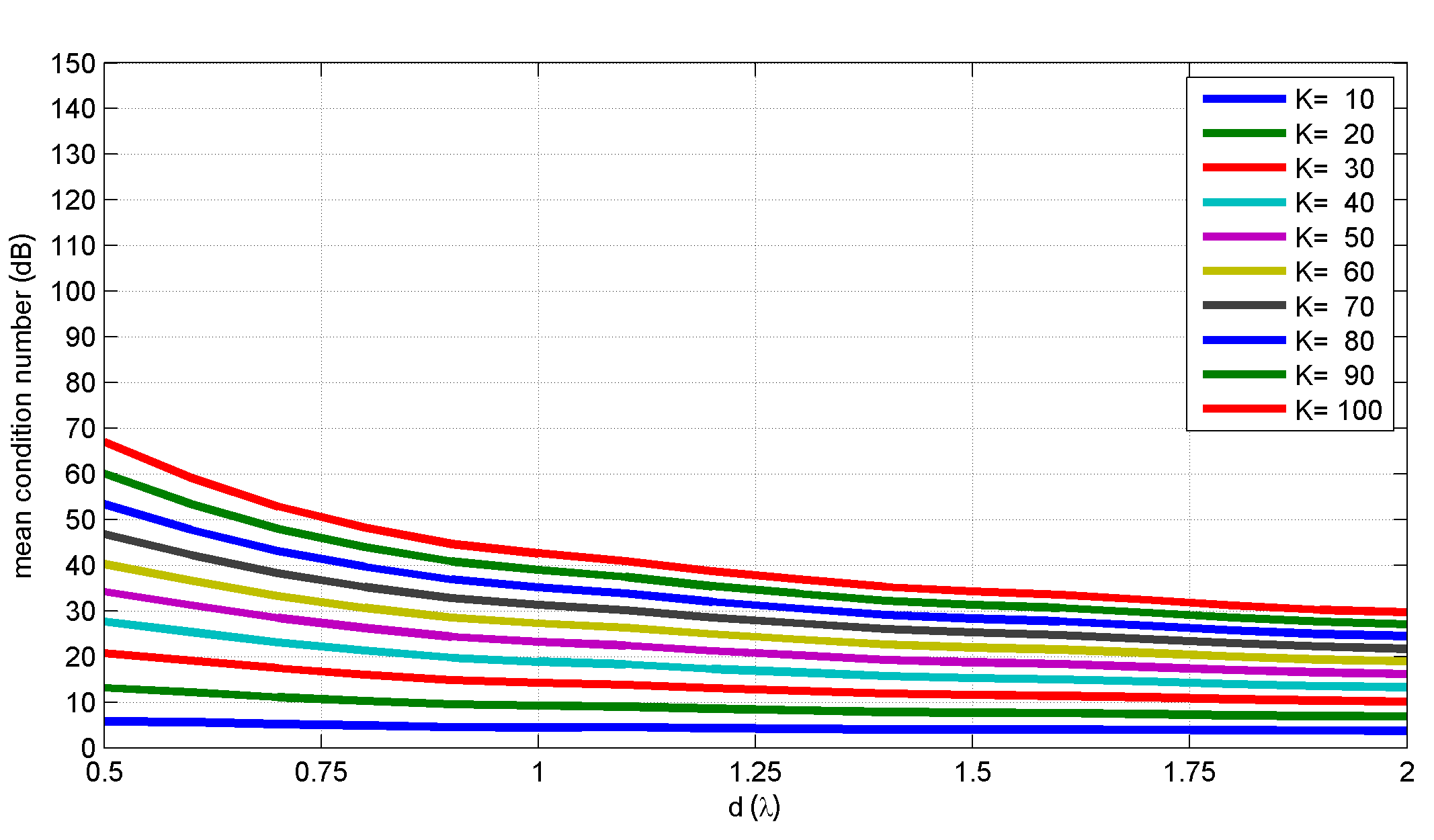}
\caption{Mean condition number of the channel matrix as a function of the average inter-element distance $d$ and the number of users $K$ for a circular array.}
\label{fig:CIRC_MeanCond_x1}
\end{figure}

In Figure \ref{fig:CIRC_MeanCond_x1}, we can see the average condition number that we can obtain for a variable inter-element distance $d$ and a variable number of users $K$. This result confirms the importance of increasing the size of the antenna array as much as possible if we are interested in improving the conditioning of the channel matrix. It is also worth noting that this result is surprisingly very similar to the result obtained with the non-equispaced antenna array of Figure~\ref{fig:MeanCond_x097}.

\begin{figure} [H]
\centering
\includegraphics[width=4in]{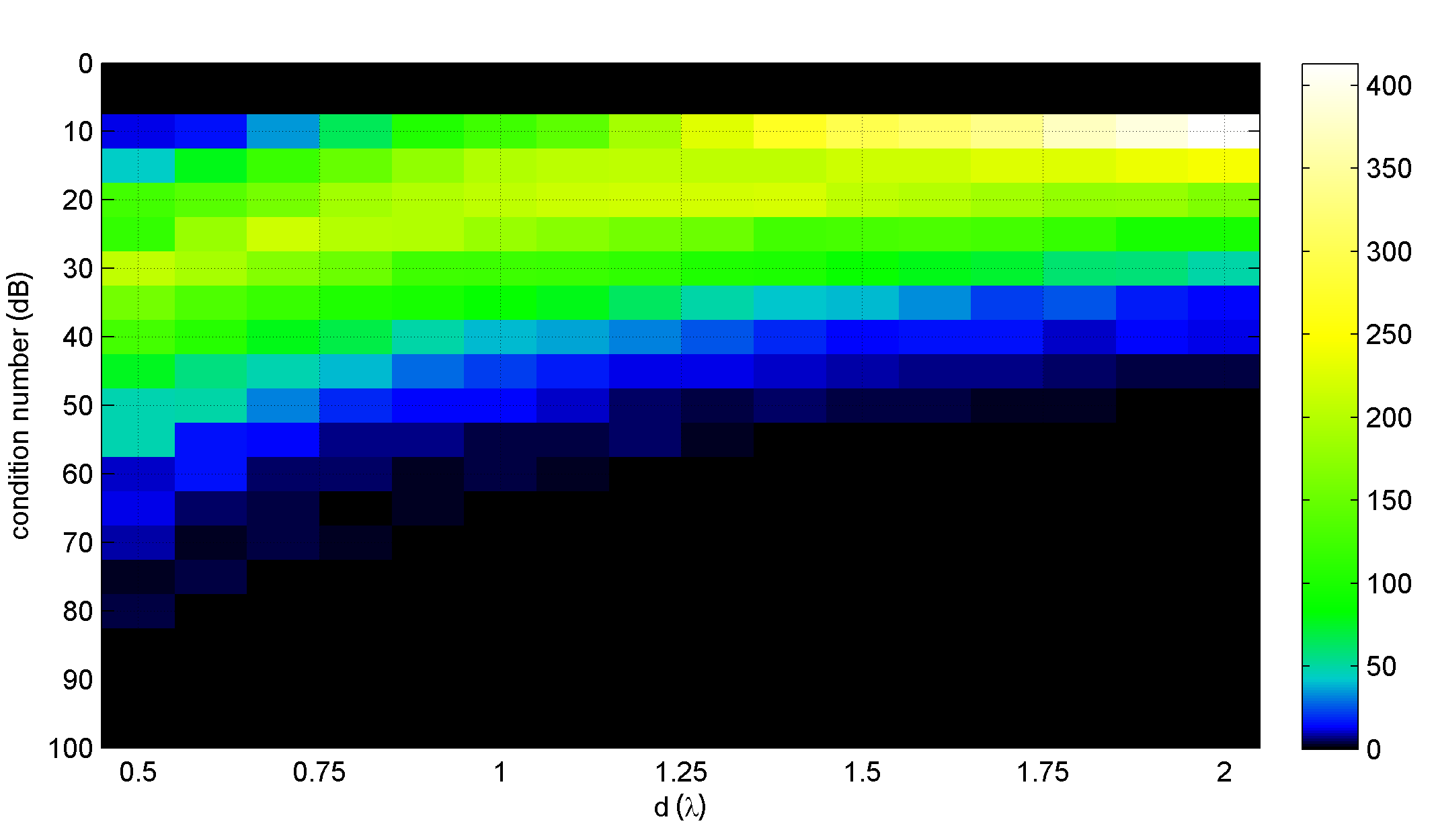}
\caption{Distribution of the condition number as a function of the average distance $d$ for the circular array with $K=50$. Each column of the matrix in the image map is a histogram of the condition~number.}
	\label{fig:CIRC_HistCond50_x1}
\end{figure}

In Figure \ref{fig:CIRC_HistCond50_x1}, we instead depict the distribution of the condition number for a variable antenna distance $d$ for the specific case $K=50$; even in this case, it is surprising to see the behavior of the condition number that is very similar to the result depicted in Figure~\ref{fig:HistCond50_x097}.

\begin{figure} [H]
\centering
\includegraphics[width=4in]{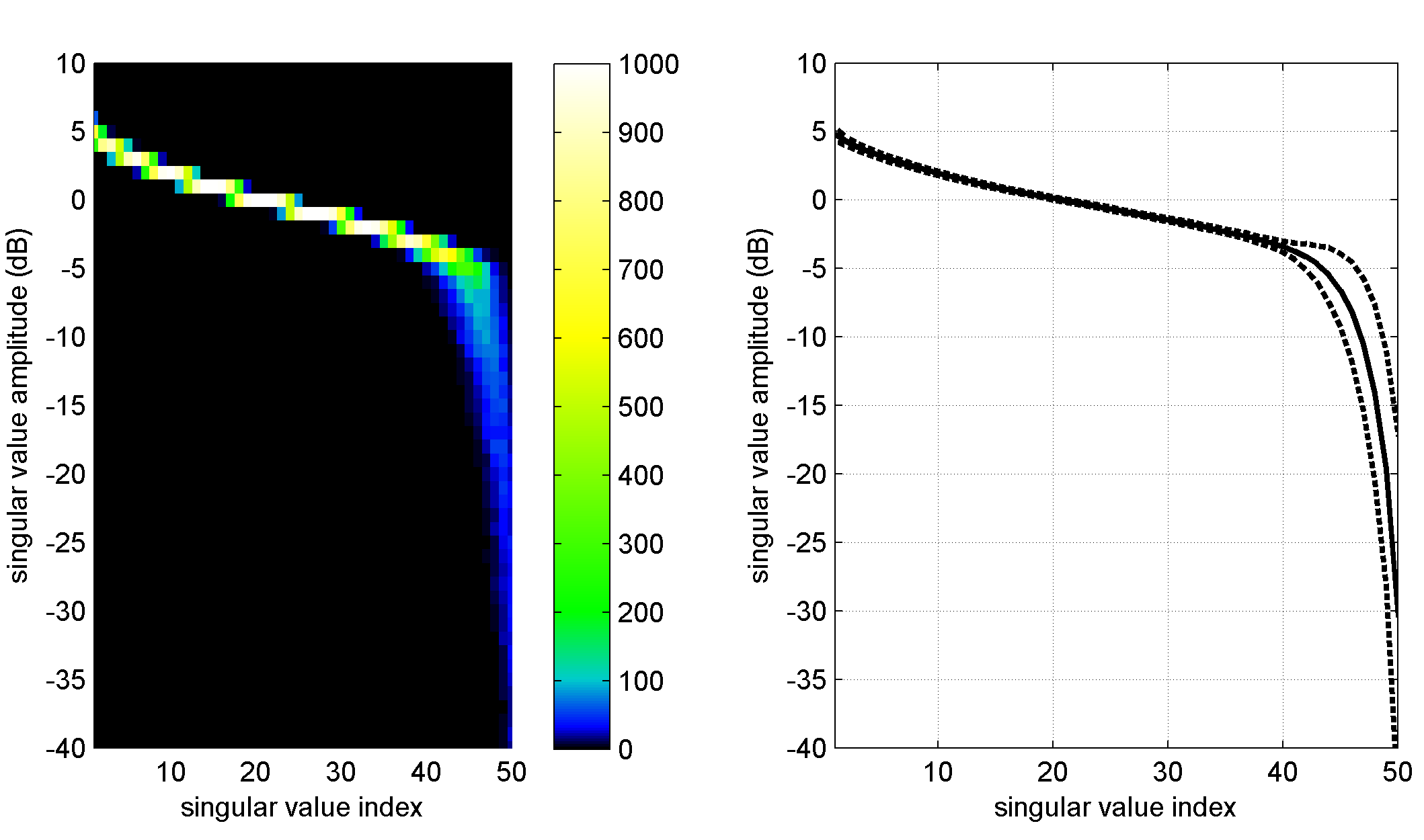}
\caption{Singular values analysis for the circular array with $d=\lambda/2$ and $K=50$. ({Left}) Image map of the distribution of each one of the 50 singular values. ({Right}) solid line, average singular values; dashed lines, average singular values $\pm$ the standard deviation of the distribution.}
\label{fig:CIRC_HistVS_d05_x1}
\end{figure}

\begin{figure} [H]
\centering
\includegraphics[width=4in]{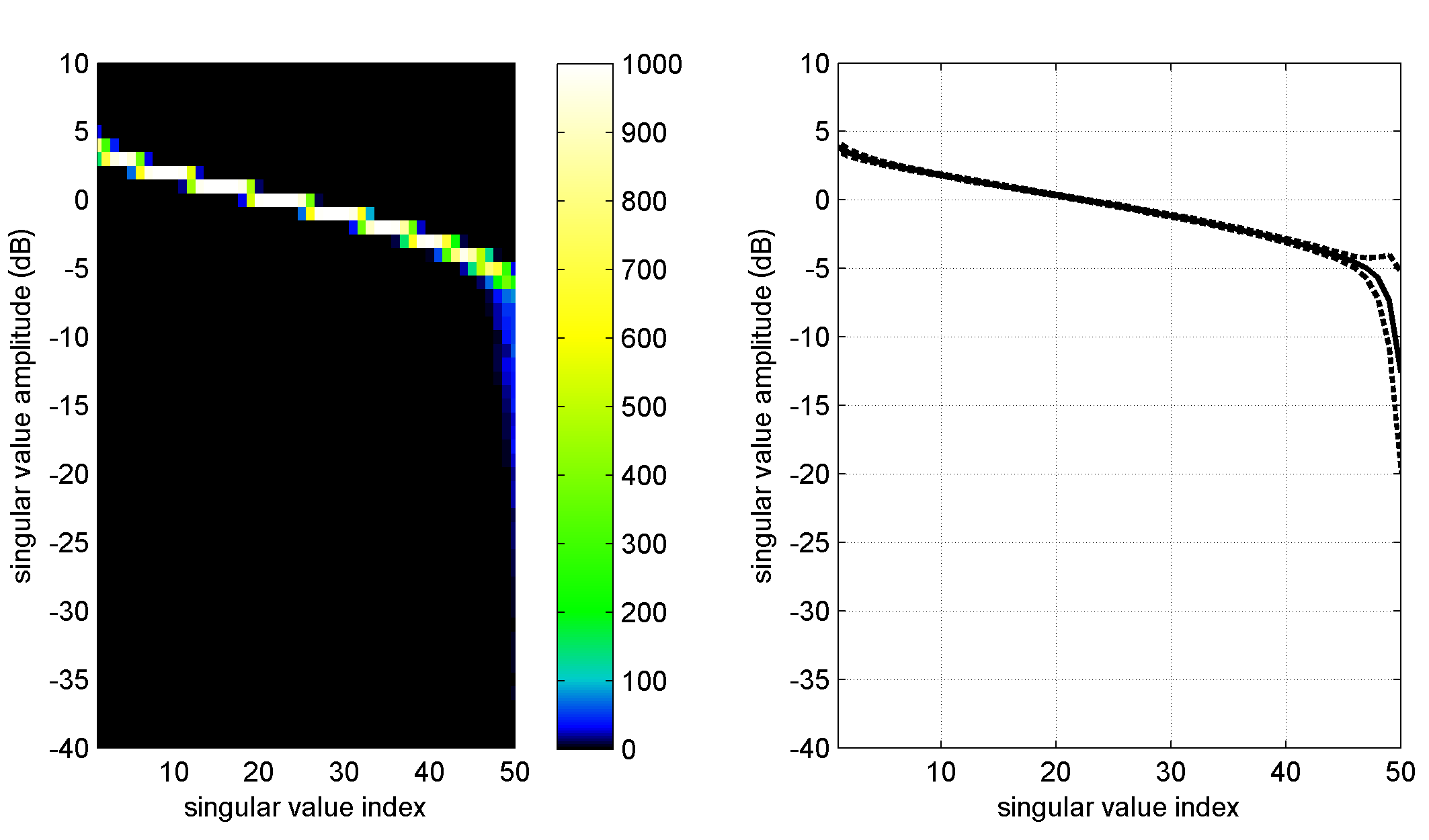}
\caption{Singular values analysis for the circular array with $d=2\lambda$ and $K=50$. ({Left}) Image map of the distribution of each one of the 50 singular values; ({right}) solid line, average singular values; dashed lines, average singular values $\pm$ the standard deviation of the distribution.}
\label{fig:CIRC_HistVS_d2_x1}
\end{figure}

Finally, in Figures \ref{fig:CIRC_HistVS_d05_x1} and \ref{fig:CIRC_HistVS_d2_x1}, we depict the distribution of the singular values for the two specific cases of $d=\lambda/2$ and $d=2\lambda$, confirming the advantage of an increased antenna array dimension.
\vspace {-6pt}
\unskip
\unskip

\section{Evaluating the Spectral Efficiency}
Up to this point, we have focused on the singular values of the channel matrix and its condition number. As stated before, this choice provides results independent of the SNR at the receiver.

A further figure of merit that is often used in MIMO system analysis is spectral efficiency. In this section, we show how the better conditioning of the channel matrix influences the spectral efficiency of massive MIMO systems. In particular, we will evaluate the average maximum achievable rate using a Zero Forcing (ZF) receiver \cite{chen2007performance}, considering a set of 1000 different scenarios for the user terminals:
\begin{equation}
C_{ZF}=\sum_{k=1}^K log\left(1+\frac{\rho}{N [(\mathbf{H}^H \mathbf{H})^{-1}]_{kk}}\right)
\end{equation}
where $\rho$ is the average SNR at the receivers and $[\mathbf{A}]_{ij}$ takes the $(i,j)$-entry of the matrix $\mathbf{A}$.

\begin{figure} [H]
\centering
\includegraphics[width=4in]{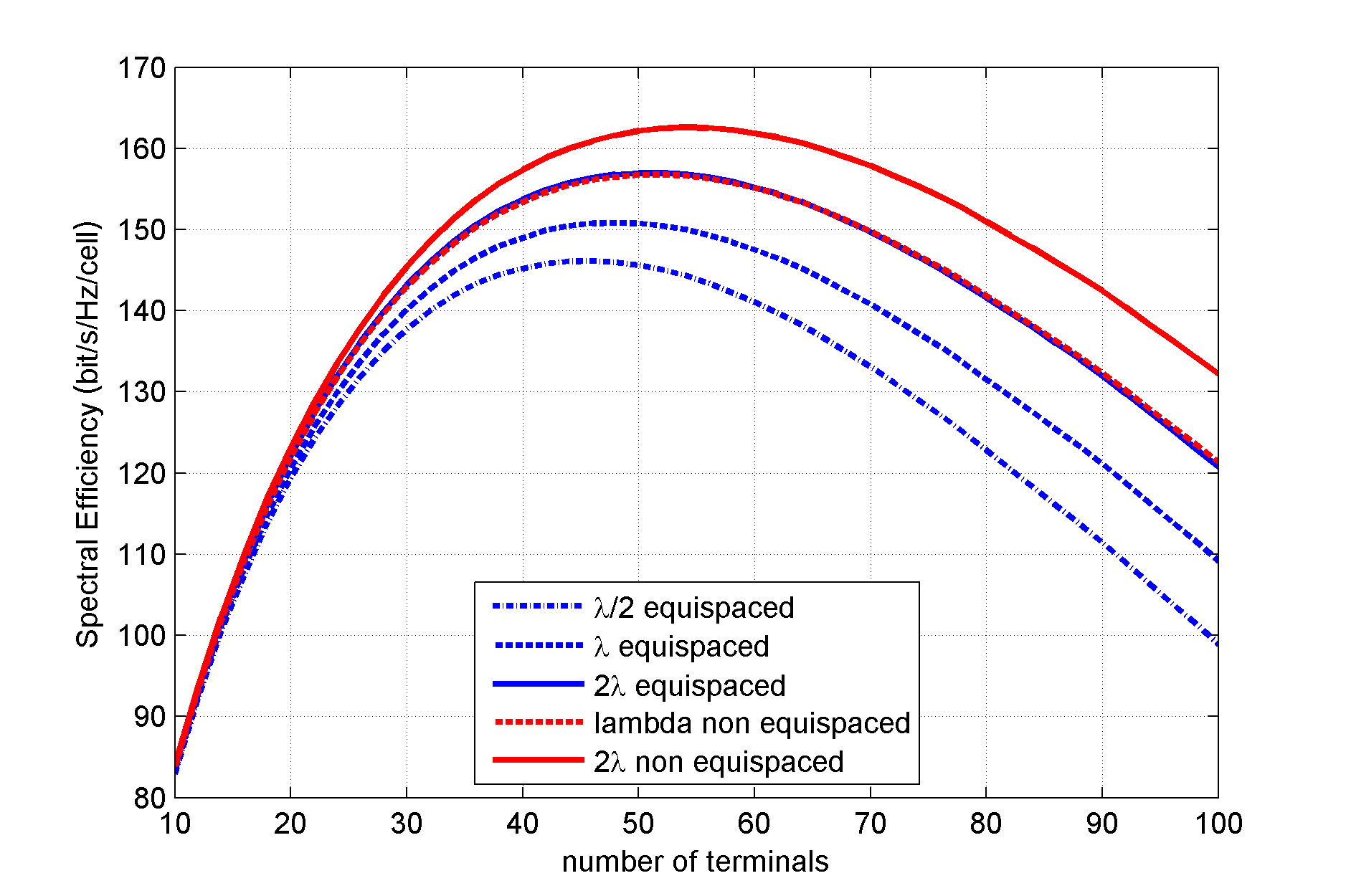}
\caption{Average maximum achievable Zero Forcing (ZF) rate as a function of the number of user terminals $K$.}
\label{fig:LinM_ZFC_varK_N200_P316}
\end{figure}

In Figure~\ref{fig:LinM_ZFC_varK_N200_P316}, we compare the performances of the equispaced array, with inter-element distances {(}$\lambda/2$, $\lambda$, $2\lambda${)}, with the performances of the non-equispaced array, calculated with a $\alpha_1=-0.03$ and average inter-element distance of {(}$\lambda$, $2\lambda${)}; the SNR at the receiver is 5 dB. It is worth noting that the maximum spectral efficiency is increased from $C=146.1$ bit/s/Hz/cell for $K=45$ of the $\lambda/2$ equispaced array to $C=162.6$ bit/s/Hz/cell for $K=55$ for the $2\lambda$ non equispaced array. 

{{We can obtain an average} spectral efficiency gain of about 11\% just positioning the radiating element of the BS in an optimal fashion.} 

\begin{figure} [H]
\centering
\includegraphics[width=4in]{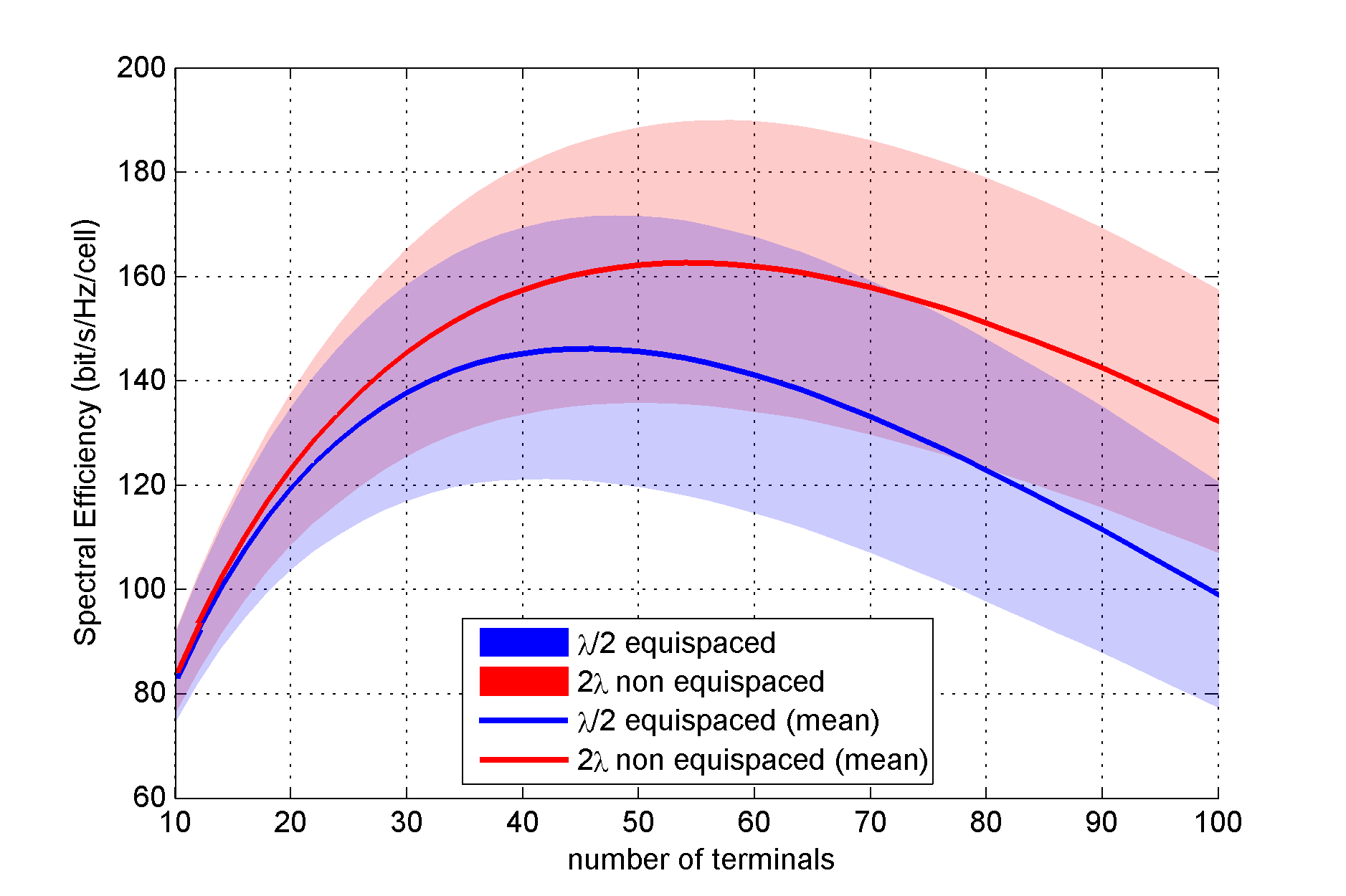}
\caption{Average maximum achievable ZF rate as a function of the number of user terminals $K$; the~areas represent the range of the standard deviation with respect to the mean.}
\label{fig:Lin_ZFC_varK_N200_P316}
\end{figure}

In Figure~\ref{fig:Lin_ZFC_varK_N200_P316}, we present the same results, but comparing only the $\lambda/2$ equispaced array and the $2\lambda$ non-equispaced array. In this plot, we also represent, by means of two colored areas, the~spreading of the spectral efficiency: the areas represent a range of a standard deviation with respect to the mean. It is interesting to compare the ``top'' border of these two areas, which gives us an indication of the performances achievable by a smart user scheduling scheme: the advantage of the non-equispaced architecture is again clear.

As a final remark, in our numerical tests, we have also considered the Dirty Paper Coding (DPC) capacity \cite{tsoulos2006mimo}, which is calculated supposing the use of the optimum receiver. In this case, the results are stable with respect to the inter-element distance and marginally depend on the use of non-equispaced~architectures. 

The reason for this fact is that only the greatest singular values of the channel matrix affect the DPC, and from the plots in Figures~\ref{fig:PROVV_HistVS_d05_x1}, \ref{fig:PROVV_HistVS_d2_x1} and \ref{fig:HistVS_d2_x097}, we can see that the distribution of the elements of the BS array influences mostly the distribution of the smallest singular values. 

\vspace {-6pt}

Actually, communication schemes that achieve the DPC capacity are very hard to implement in the massive MIMO framework, where linear processing (like ZF) is preferred for computing reasons, and for this reason, we decided to show only the ZF performances.

\section{Some Considerations and Antenna Array Design Guidelines}
In this section, we would like to sum up some design consideration that can be inferred by the shown results.

All of the simulations have been performed focusing on arrays with $N=200$ radiating elements, but we have also performed other simulations, with a different number of radiators, confirming the advantages of sparse arrays for massive MIMO applications.

We have indeed chosen a very simple antenna model for the simulations (an array of isotropic elements without mutual coupling), but in the examples reported in the paper, the inter-element spacing is always larger than $0.5$ wavelengths, and in some cases, it is much larger. For such inter-element distances, it is well known that mutual coupling usually plays a minor role. Anyway, we~have performed some numerical tests including the mutual coupling ({not reported} here for lack of space), and we found that there was no significant modification of the distribution of the singular values. Similarly, the element pattern and the polarization will provide a different amplification of the signal received by the terminals, but the overall impact on the singular values distribution would be small. For~these reasons, we preferred the use of the architecture of isotropic elements without mutual coupling, providing more easily repeatable results by interested readers.

We would also like to underline that the proposed increase in size of the array can have an effect on the overall cost of the antenna system, but this increment would not be significant when dealing with mm-wave frequencies. It is indeed true that the optimization of the BS antenna that we propose has to be realized only when setting up the system; no optimizations/antenna selections are required during operation.

On the contrary, the use of non-equispaced architectures, instead of equispaced ones, should not have, in general, a direct impact on the overall cost or complexity of the communication system. The~only relevant effect that should be taken into account in an architecture trade-off is the difficulty of employing modular BS arrays (i.e., antenna arrays divided in equal modules of a fixed number of elements): in this case, it could be advantageous to design non-equispaced modules for the BS array and placing them in a non-regular fashion, but the discussion of this possibility goes beyond the aims of this paper.

A final consideration is now needed. In the numerical tests shown, we have considered the LoS component only; even if this is a reasonable assumption, we could argue if the shown results hold also when a non-LoS propagation component is present. Some preliminary simulations show that the advantage of using larger, non-equispaced arrays is still present, but~smaller. Anyway, the analysis of this case deserves particular attention and will be the main topic of a forthcoming work.

Summing up, we would like to list some rules-of-thumb for designing massive MIMO arrays that work in LoS propagation.

\begin{enumerate}
	\item The antenna array geometry needs to be matched to the expected user terminal displacement. For sectoral coverage, a linear array is preferable, while for circular coverage, a circular antenna works better, since a linear array is not capable of discriminating between two terminals in specular positions with respect to the array.
	\item Given the number of radiating elements, and hence transceivers, that can be used in the MIMO system, the BS antenna array should be as large as possible. 
	\item Once the maximum dimension is used, in a linear array, the radiating elements should be possibly arranged in a non-equispaced fashion, by using for instance the positions provided by Equation~(\ref{eq:TchSparse}).
\end{enumerate}

\section{Conclusions}
In this paper, we have discussed the impact of the array on the performances of a massive MIMO~system. 

In particular, we have analyzed the LoS case, showing that increasing the size of the array can be very beneficial to the condition number of the MU-MIMO channel matrix: since the BS antenna is working in the near-field region of the array, a larger element distance allows one to exploit the spherical shape of the wavefronts. 

We have also shown that using a non-equispaced array, we are able to further improve the resulting condition number, since by getting rid of the grating lobes, we are able to achieve a significant lowering of the average condition number also for small increases of the antenna array dimension. Such an improvement in the conditioning of the channel matrix results in an improvement in the spectral efficiency of the massive MIMO systems when using linear processing of the signals.

We have also analyzed the results achievable by means of circular arrays and a full circle coverage, finding results perfectly aligned with the ones of the non-equispaced linear array.

It is worth underlining that all of the performed analyses have been done considering a fixed BS complexity. It is trivial to understand that with a proper sparse array design, we could reduce the number of antennas at the BS, achieving the same average conditioning of the channel matrix for a desired number of user terminals. 

It is fundamental to underline that the improvement that can be achieved by optimizing the antenna array geometry has a minimum impact on the overall system cost and does not increase the computational effort of the communication schemes in any way.

As a development of the present work, we are working on the design of more complex planar and conformal antenna array geometries; we are also working towards the use of a ray-tracing simulator, in~order to check the achieved results when the line of sight is the dominant, but not the only, propagation term present. Finally, we will investigate the effect of a joint optimization of the antenna array when using a smart user scheduling approach.

\vspace {12pt}
\authorcontributions{{D.P.
 conceived of the approach and wrote the numerical code to perform the simulations. M.D.M, G.P. and F.S. helped with the discussion and comparison of the results. D.P., M.D.M. and F.S. wrote~the~paper.}}
\vspace {6pt}

\conflictsofinterest{The authors declare no conflict of interest.} 

\bibliographystyle{mdpi}

\bibliography{biblio_MASSIVE_arxiv}

\begin{thebibliography}{-------}
\providecommand{\natexlab}[1]{#1}

\bibitem[Hoydis \em{et~al.}(2011)Hoydis, Ten~Brink, and
  Debbah]{hoydis2011massive}
Hoydis, J.; Ten~Brink, S.; Debbah, M.
\newblock Massive MIMO: How many antennas do we need?
\newblock  Communication, Control, and Computing (Allerton), 2011 49th Annual
  Allerton Conference on. IEEE,  2011, pp. 545--550.

\bibitem[Huh \em{et~al.}(2012)Huh, Caire, Papadopoulos, and
  Ramprashad]{huh2012achieving}
Huh, H.; Caire, G.; Papadopoulos, H.C.; Ramprashad, S.A.
\newblock Achieving massive MIMO spectral efficiency with a not-so-large number
  of antennas.
\newblock {\em IEEE Transactions on Wireless Communications} {\bf 2012}, {\em
  11},~3226--3239.

\bibitem[Rusek \em{et~al.}(2013)Rusek, Persson, Lau, Larsson, Marzetta, Edfors,
  and Tufvesson]{rusek2013scaling}
Rusek, F.; Persson, D.; Lau, B.K.; Larsson, E.G.; Marzetta, T.L.; Edfors, O.;
  Tufvesson, F.
\newblock Scaling up MIMO: Opportunities and challenges with very large arrays.
\newblock {\em IEEE Signal Processing Magazine} {\bf 2013}, {\em 30},~40--60.

\bibitem[Hoydis \em{et~al.}(2013)Hoydis, Hosseini, Brink, and
  Debbah]{hoydis2013making}
Hoydis, J.; Hosseini, K.; Brink, S.t.; Debbah, M.
\newblock Making smart use of excess antennas: Massive MIMO, small cells, and
  TDD.
\newblock {\em Bell Labs Technical Journal} {\bf 2013}, {\em 18},~5--21.

\bibitem[Nam \em{et~al.}(2013)Nam, Ng, Sayana, Li, Zhang, Kim, and
  Lee]{nam2013full}
Nam, Y.H.; Ng, B.L.; Sayana, K.; Li, Y.; Zhang, J.; Kim, Y.; Lee, J.
\newblock Full-dimension MIMO (FD-MIMO) for next generation cellular
  technology.
\newblock {\em IEEE Communications Magazine} {\bf 2013}, {\em 51},~172--179.

\bibitem[Hoydis \em{et~al.}(2013)Hoydis, Ten~Brink, and
  Debbah]{hoydis2013massive}
Hoydis, J.; Ten~Brink, S.; Debbah, M.
\newblock Massive MIMO in the UL/DL of cellular networks: How many antennas do
  we need?
\newblock {\em IEEE Journal on selected Areas in Communications} {\bf 2013},
  {\em 31},~160--171.

\bibitem[Lu \em{et~al.}(2014)Lu, Li, Swindlehurst, Ashikhmin, and
  Zhang]{lu2014overview}
Lu, L.; Li, G.Y.; Swindlehurst, A.L.; Ashikhmin, A.; Zhang, R.
\newblock An overview of massive MIMO: Benefits and challenges.
\newblock {\em IEEE journal of selected topics in signal processing} {\bf
  2014}, {\em 8},~742--758.

\bibitem[Boccardi \em{et~al.}(2014)Boccardi, Heath, Lozano, Marzetta, and
  Popovski]{boccardi2014five}
Boccardi, F.; Heath, R.W.; Lozano, A.; Marzetta, T.L.; Popovski, P.
\newblock Five disruptive technology directions for 5G.
\newblock {\em IEEE Communications Magazine} {\bf 2014}, {\em 52},~74--80.

\bibitem[Jungnickel \em{et~al.}(2014)Jungnickel, Manolakis, Zirwas, Panzner,
  Braun, Lossow, Sternad, Apelfrojd, and Svensson]{jungnickel2014role}
Jungnickel, V.; Manolakis, K.; Zirwas, W.; Panzner, B.; Braun, V.; Lossow, M.;
  Sternad, M.; Apelfrojd, R.; Svensson, T.
\newblock The role of small cells, coordinated multipoint, and massive MIMO in
  5G.
\newblock {\em IEEE Communications Magazine} {\bf 2014}, {\em 52},~44--51.

\bibitem[Larsson \em{et~al.}(2014)Larsson, Edfors, Tufvesson, and
  Marzetta]{larsson2014massive}
Larsson, E.G.; Edfors, O.; Tufvesson, F.; Marzetta, T.L.
\newblock Massive MIMO for next generation wireless systems.
\newblock {\em IEEE Communications Magazine} {\bf 2014}, {\em 52},~186--195.

\bibitem[Bj{\"o}rnson \em{et~al.}(2015)Bj{\"o}rnson, Sanguinetti, Hoydis, and
  Debbah]{bjornson2015optimal}
Bj{\"o}rnson, E.; Sanguinetti, L.; Hoydis, J.; Debbah, M.
\newblock Optimal design of energy-efficient multi-user MIMO systems: Is
  massive MIMO the answer?
\newblock {\em IEEE Transactions on Wireless Communications} {\bf 2015}, {\em
  14},~3059--3075.

\bibitem[Bj{\"o}rnson \em{et~al.}(2016)Bj{\"o}rnson, Larsson, and
  Marzetta]{bjornson2016massive}
Bj{\"o}rnson, E.; Larsson, E.G.; Marzetta, T.L.
\newblock Massive MIMO: Ten myths and one critical question.
\newblock {\em IEEE Communications Magazine} {\bf 2016}, {\em 54},~114--123.

\bibitem[Swindlehurst \em{et~al.}(2014)Swindlehurst, Ayanoglu, Heydari, and
  Capolino]{swindlehurst2014millimeter}
Swindlehurst, A.L.; Ayanoglu, E.; Heydari, P.; Capolino, F.
\newblock Millimeter-wave massive MIMO: the next wireless revolution?
\newblock {\em IEEE Communications Magazine} {\bf 2014}, {\em 52},~56--62.

\bibitem[Lee \em{et~al.}(2016)Lee, Sung, and Seo]{lee2016randomly}
Lee, G.; Sung, Y.; Seo, J.
\newblock Randomly-directional beamforming in millimeter-wave multiuser MISO
  downlink.
\newblock {\em IEEE Transactions on Wireless Communications} {\bf 2016}, {\em
  15},~1086--1100.

\bibitem[Puglielli \em{et~al.}(2016)Puglielli, Townley, LaCaille,
  Milovanovi{\'c}, Lu, Trotskovsky, Whitcombe, Narevsky, Wright, Courtade,
  et~al.]{puglielli2016design}
Puglielli, A.; Townley, A.; LaCaille, G.; Milovanovi{\'c}, V.; Lu, P.;
  Trotskovsky, K.; Whitcombe, A.; Narevsky, N.; Wright, G.; Courtade, T.;
  others.
\newblock Design of energy-and cost-efficient massive MIMO arrays.
\newblock {\em Proceedings of the IEEE} {\bf 2016}, {\em 104},~586--606.

\bibitem[Sarkar \em{et~al.}(2003)Sarkar, Ji, Kim, Medouri, and
  Salazar-Palma]{sarkar2003survey}
Sarkar, T.K.; Ji, Z.; Kim, K.; Medouri, A.; Salazar-Palma, M.
\newblock A survey of various propagation models for mobile communication.
\newblock {\em IEEE Antennas and propagation Magazine} {\bf 2003}, {\em
  45},~51--82.

\bibitem[Medbo \em{et~al.}(2014)Medbo, B{\"o}rner, Haneda, Hovinen, Imai,
  J{\"a}rvelainen, J{\"a}ms{\"a}, Karttunen, Kusume, Kyr{\"o}l{\"a}inen,
  et~al.]{medbo2014channel}
Medbo, J.; B{\"o}rner, K.; Haneda, K.; Hovinen, V.; Imai, T.; J{\"a}rvelainen,
  J.; J{\"a}ms{\"a}, T.; Karttunen, A.; Kusume, K.; Kyr{\"o}l{\"a}inen, J.;
  others.
\newblock Channel modelling for the fifth generation mobile communications.
\newblock  Antennas and Propagation (EuCAP), 2014 8th European Conference on.
  IEEE,  2014, pp. 219--223.

\bibitem[Ngo \em{et~al.}(2014)Ngo, Larsson, and Marzetta]{ngo2014aspects}
Ngo, H.Q.; Larsson, E.G.; Marzetta, T.L.
\newblock Aspects of favorable propagation in massive MIMO.
\newblock  2014 22nd European Signal Processing Conference (EUSIPCO). IEEE,
  2014, pp. 76--80.

\bibitem[Migliore(2014)]{migliore2014some}
Migliore, M.D.
\newblock Some Electromagnetic Limitations on the Number of Users in MU-MIMO
  Communication Systems.
\newblock {\em IEEE Antennas and Wireless Propagation Letters} {\bf 2014}, {\em
  13},~181--184.

\bibitem[Gao \em{et~al.}(2015)Gao, Edfors, Rusek, and
  Tufvesson]{gao2015massive}
Gao, X.; Edfors, O.; Rusek, F.; Tufvesson, F.
\newblock Massive MIMO performance evaluation based on measured propagation
  data.
\newblock {\em IEEE Transactions on Wireless Communications} {\bf 2015}, {\em
  14},~3899--3911.

\bibitem[Lee \em{et~al.}(2015)Lee, Sung, and Kountouris]{lee2015performance}
Lee, G.; Sung, Y.; Kountouris, M.
\newblock On the performance of randomly directional beamforming between
  line-of-sight and rich scattering channels.
\newblock  Signal Processing Advances in Wireless Communications (SPAWC), 2015
  IEEE 16th International Workshop on. IEEE,  2015, pp. 141--145.

\bibitem[Liu \em{et~al.}(2016)Liu, Matolak, Tao, Li, Ai, and
  Chen]{liu2016channel}
Liu, L.; Matolak, D.W.; Tao, C.; Li, Y.; Ai, B.; Chen, H.
\newblock Channel capacity investigation of a linear massive MIMO system using
  spherical wave model in LOS scenarios.
\newblock {\em Science China Information Sciences} {\bf 2016}, {\em 59},~1--15.

\bibitem[Chandhar \em{et~al.}(2016)Chandhar, Danev, and
  Larsson]{chandhar2016ergodic}
Chandhar, P.; Danev, D.; Larsson, E.G.
\newblock On ergodic rates and optimal array geometry in line-of-sight massive
  MIMO.
\newblock  Signal Processing Advances in Wireless Communications (SPAWC), 2016
  IEEE 17th International Workshop on. IEEE,  2016, pp. 1--6.

\bibitem[Liu \em{et~al.}(2016)Liu, Matolak, Tao, Lu, Ai, and
  Chen]{liu2016geometry}
Liu, L.; Matolak, D.W.; Tao, C.; Lu, Y.; Ai, B.; Chen, H.
\newblock Geometry based large scale attenuation over linear massive MIMO
  systems.
\newblock  Antennas and Propagation (EuCAP), 2016 10th European Conference on.
  IEEE,  2016, pp. 1--5.

\bibitem[Zhou \em{et~al.}(2015)Zhou, Gao, Fang, and Chen]{zhou2015spherical}
Zhou, Z.; Gao, X.; Fang, J.; Chen, Z.
\newblock Spherical wave channel and analysis for large linear array in LoS
  conditions.
\newblock  Globecom Workshops (GC Wkshps), 2015 IEEE. IEEE,  2015, pp. 1--6.

\bibitem[Chen \em{et~al.}(2016)Chen, Volskiy, Chiumento, Van~der Perre,
  Vandenbosch, and Pollin]{chen2016exploration}
Chen, C.M.; Volskiy, V.; Chiumento, A.; Van~der Perre, L.; Vandenbosch, G.A.;
  Pollin, S.
\newblock Exploration of User Separation Capabilities by Distributed Large
  Antenna Arrays.
\newblock  Globecom Workshops (GC Wkshps), 2016 IEEE. IEEE,  2016, pp. 1--6.

\bibitem[Liu \em{et~al.}(2016)Liu, Matolak, Tao, Li, and Chen]{liu2016benefits}
Liu, L.; Matolak, D.W.; Tao, C.; Li, Y.; Chen, H.
\newblock The Benefits of Large-Scale Attenuation over the Antenna Array in
  Massive MIMO Systems.
\newblock  Vehicular Technology Conference (VTC-Fall), 2016 IEEE 84th. IEEE,
  2016, pp. 1--5.

\bibitem[Masouros and Matthaiou(2015)]{masouros2015space}
Masouros, C.; Matthaiou, M.
\newblock Space-constrained massive MIMO: Hitting the wall of favorable
  propagation.
\newblock {\em IEEE Communications Letters} {\bf 2015}, {\em 19},~771--774.

\bibitem[Golub and Van~Loan(2012)]{golub2012matrix}
Golub, G.H.; Van~Loan, C.F.
\newblock {\em Matrix computations}; Vol.~3, JHU Press,  2012.

\bibitem[Tsoulos(2006)]{tsoulos2006mimo}
Tsoulos, G.
\newblock {\em MIMO system technology for wireless communications}; CRC press,
  2006.

\bibitem[Art{\'e}s \em{et~al.}(2003)Art{\'e}s, Seethaler, and
  Hlawatsch]{artes2003efficient}
Art{\'e}s, H.; Seethaler, D.; Hlawatsch, F.
\newblock Efficient detection algorithms for MIMO channels: A geometrical
  approach to approximate ML detection.
\newblock {\em IEEE transactions on signal processing} {\bf 2003}, {\em
  51},~2808--2820.

\bibitem[Maurer \em{et~al.}(2007)Maurer, Matz, and Seethaler]{maurer2007low}
Maurer, J.; Matz, G.; Seethaler, D.
\newblock Low-complexity and full-diversity MIMO detection based on condition
  number thresholding.
\newblock  Acoustics, Speech and Signal Processing, 2007. ICASSP 2007. IEEE
  International Conference on. IEEE,  2007, Vol.~3, pp. III--61.

\bibitem[Mohammed \em{et~al.}(2011)Mohammed, Viterbo, Hong, and
  Chockalingam]{mohammed2011mimo}
Mohammed, S.K.; Viterbo, E.; Hong, Y.; Chockalingam, A.
\newblock MIMO precoding with X-and Y-codes.
\newblock {\em IEEE Transactions on Information Theory} {\bf 2011}, {\em
  57},~3542--3566.

\bibitem[Yang and Marzetta(2013)]{yang2013performance}
Yang, H.; Marzetta, T.L.
\newblock Performance of conjugate and zero-forcing beamforming in large-scale
  antenna systems.
\newblock {\em IEEE Journal on Selected Areas in Communications} {\bf 2013},
  {\em 31},~172--179.

\bibitem[Mailloux(2005)]{mailloux2005phased}
Mailloux, R.J.
\newblock {\em Phased array antenna handbook}; Vol.~2, Artech House Boston,
  2005.

\bibitem[Donelli \em{et~al.}(2004)Donelli, Caorsi, DeNatale, Pastorino, and
  Massa]{donelli2004linear}
Donelli, M.; Caorsi, S.; DeNatale, F.; Pastorino, M.; Massa, A.
\newblock Linear antenna synthesis with a hybrid genetic algorithm.
\newblock {\em Progress In Electromagnetics Research} {\bf 2004}, {\em
  49},~1--22.

\bibitem[Massa \em{et~al.}(2004)Massa, Donelli, De~Natale, Caorsi, and
  Lommi]{massa2004planar}
Massa, A.; Donelli, M.; De~Natale, F.G.; Caorsi, S.; Lommi, A.
\newblock Planar antenna array control with genetic algorithms and adaptive
  array theory.
\newblock {\em IEEE Transactions on Antennas and Propagation} {\bf 2004}, {\em
  52},~2919--2924.

\bibitem[Liu \em{et~al.}(2008)Liu, Nie, and Liu]{liu2008reducing}
Liu, Y.; Nie, Z.; Liu, Q.H.
\newblock Reducing the number of elements in a linear antenna array by the
  matrix pencil method.
\newblock {\em IEEE Transactions on Antennas and Propagation} {\bf 2008}, {\em
  56},~2955--2962.

\bibitem[Oliveri \em{et~al.}(2009)Oliveri, Donelli, and
  Massa]{oliveri2009linear}
Oliveri, G.; Donelli, M.; Massa, A.
\newblock Linear array thinning exploiting almost difference sets.
\newblock {\em IEEE Transactions on Antennas and Propagation} {\bf 2009}, {\em
  57},~3800--3812.

\bibitem[Poli \em{et~al.}(2013)Poli, Rocca, Salucci, and
  Massa]{poli2013reconfigurable}
Poli, L.; Rocca, P.; Salucci, M.; Massa, A.
\newblock Reconfigurable thinning for the adaptive control of linear arrays.
\newblock {\em IEEE Transactions on Antennas and Propagation} {\bf 2013}, {\em
  61},~5068--5077.

\bibitem[Oliveri \em{et~al.}(2014)Oliveri, Bekele, Robol, and
  Massa]{oliveri2014sparsening}
Oliveri, G.; Bekele, E.T.; Robol, F.; Massa, A.
\newblock Sparsening conformal arrays through a versatile $ BCS $-based method.
\newblock {\em IEEE Transactions on Antennas and Propagation} {\bf 2014}, {\em
  62},~1681--1689.

\bibitem[Bucci \em{et~al.}(2015)Bucci, Perna, and Pinchera]{bucci2015synthesis}
Bucci, O.M.; Perna, S.; Pinchera, D.
\newblock Synthesis of isophoric sparse arrays allowing zoomable beams and
  arbitrary coverage in satellite communications.
\newblock {\em IEEE Transactions on Antennas and Propagation} {\bf 2015}, {\em
  63},~1445--1457.

\bibitem[D'Urso \em{et~al.}(2016)D'Urso, Prisco, and Tumolo]{d2016maximally}
D'Urso, M.; Prisco, G.; Tumolo, R.M.
\newblock Maximally Sparse, Steerable, and Nonsuperdirective Array Antennas via
  Convex Optimizations.
\newblock {\em IEEE Transactions on Antennas and Propagation} {\bf 2016}, {\em
  64},~3840--3849.

\bibitem[Pinchera \em{et~al.}(2016)Pinchera, Migliore, Lucido, Schettino, and
  Panariello]{pinchera2016compressive}
Pinchera, D.; Migliore, M.D.; Lucido, M.; Schettino, F.; Panariello, G.
\newblock A Compressive-Sensing Inspired Alternate Projection Algorithm for
  Sparse Array Synthesis.
\newblock {\em Electronics} {\bf 2016}, {\em 6},~3.

\bibitem[Rocca \em{et~al.}(2016)Rocca, Oliveri, Mailloux, and
  Massa]{rocca2016unconventional}
Rocca, P.; Oliveri, G.; Mailloux, R.J.; Massa, A.
\newblock Unconventional phased array architectures and design
  methodologies—A review.
\newblock {\em Proceedings of the IEEE} {\bf 2016}, {\em 104},~544--560.

\bibitem[Pinchera and Migliore(2016)]{pinchera2016comparison}
Pinchera, D.; Migliore, M.D.
\newblock Comparison guidelines and benchmark procedure for sparse array
  synthesis.
\newblock {\em Progress In Electromagnetics Research M} {\bf 2016}, {\em
  52},~129--139.

\bibitem[Bucci \em{et~al.}(2017)Bucci, Perna, and
  Pinchera]{bucci2017interleaved}
Bucci, O.M.; Perna, S.; Pinchera, D.
\newblock Interleaved Isophoric Sparse Arrays for the Radiation of Steerable
  and Switchable Beams in Satellite Communications.
\newblock {\em IEEE Transactions on Antennas and Propagation} {\bf 2017}, {\em
  65},~1163--1173.

\bibitem[Nemirovskii \em{et~al.}(1983)Nemirovskii, Yudin, and
  Dawson]{nemirovskii1983problem}
Nemirovskii, A.; Yudin, D.B.; Dawson, E.R.
\newblock Problem complexity and method efficiency in optimization {\bf 1983}.

\bibitem[Chen and Wang(2007)]{chen2007performance}
Chen, C.J.; Wang, L.C.
\newblock Performance analysis of scheduling in multiuser MIMO systems with
  zero-forcing receivers.
\newblock {\em IEEE Journal on Selected Areas in Communications} {\bf 2007},
  {\em 25}.

\end{thebibliography}

\end{document}